\documentclass[12pt]{article}
\usepackage{amsfonts,amsthm,amsmath,amssymb,upgreek,bm}
\usepackage{hyperref}
\usepackage[paper=letterpaper,margin=.85in]{geometry}
\usepackage{graphicx}
\usepackage{units}
\usepackage{float}
\usepackage[export]{adjustbox}
\usepackage{xcolor}
\usepackage[shortlabels]{enumitem}
\usepackage[mathscr]{euscript}
\usepackage[normalem]{ulem}
\usepackage{bm}
\usepackage{tikz}
\usepackage{braket}
\usetikzlibrary{decorations.pathmorphing}
\tikzset{snake it/.style={decorate, decoration=snake}}

\newcommand{\be}{\begin{equation}}
\newcommand{\ee}{\end{equation}}
\newcommand{\bea}{\begin{eqnarray}}
\newcommand{\eea}{\end{eqnarray}}
\renewcommand{\i}{\mathrm{i}}
\renewcommand{\d}{\mathrm{d}}

\newcommand{\tm}{\tilde{t}}
\newcommand{\xm}{\tilde{x}}
\newcommand{\LambdaL}{\Lambda}
\newcommand{\ppole}{{\sf P_{pole}}}

\newcommand{\h}{{\sf h}}

\newcommand{\DrawBracket}[3]{
  \begin{scope}[shift={(#1,#2)},rotate = #3]
  \draw[very thick, rounded corners = .2mm] (-.2,+.2) -- (-.3,+.2) -- (-.3,-.2) -- (-.2,-.2);
  \draw[very thick, rounded corners = .2mm] (+.2,+.2) -- (+.3,+.2) -- (+.3,-.2) -- (+.2,-.2);
  \end{scope}
}

\newcommand{\lr}{\langle}

\newcommand{\rr}{\rangle}

\def\ri{\right}
\newcommand{\ha}{\frac{1}{2}}
\newcommand{\nn}{\nonumber\\}

\numberwithin{equation}{section}

\begin{document}
\thispagestyle{empty}

\vspace*{2.5cm}
\begin{center}

{\bf {\LARGE Scramblon loops}}\\

\begin{center}

\vspace{1cm}

{\bf Douglas Stanford, Shreya Vardhan, and Shunyu Yao}\\
 \bigskip \rm

\bigskip 

Stanford Institute for Theoretical Physics,\\Stanford University, Stanford, CA 94305

\rm
  \end{center}

\vspace{2.5cm}
{\bf Abstract}
\end{center}
\begin{quotation}
\noindent

In large $N$ chaotic quantum systems, the butterfly effect is mediated by a collective field mode known as the ``scramblon.'' We study self-interactions of the scramblon in variants of the Sachdev-Ye-Kitaev model. In spatially extended versions of the model and for large spatial separation, fluctuations described by loop diagrams can invalidate the single-scramblon approximation well before its contribution to out-of-time-order correlators becomes of order one. We find a qualitative difference between an incoherent regime at high temperaure (or in a Brownian version of the model) and a coherent regime at low temperature.

\end{quotation}

\setcounter{page}{0}
\setcounter{tocdepth}{2}
\setcounter{footnote}{0}
\newpage

\parskip 0.1in
 
\setcounter{page}{2}
\tableofcontents

\section{Introduction}

The out-of-time-ordered correlator (OTOC) is a probe of quantum many-body chaos that has been extensively studied in the context of both quantum gravity and condensed matter physics, starting with \cite{larkin1969quasiclassical,Almheiri:2013hfa,Shenker:2013pqa,kitaevfundamental} (see \cite{xu_review} for a review). The OTOC reveals the time evolution of simple few-body operators into complicated non-local operators.

In large $N$ systems such as holographic CFTs and generalized SYK models, this operator growth is reflected at leading order in $1/N$ by the exponential growth of the connected OTOC with time, together with an exponential decay in space. This exponential growth can be understood as the propagator for a ``scramblon'' collective field that appears when one studies such models on the type of time contour needed to compute an OTOC (this is a double Keldysh-Schwinger contour, with two time-folds). In the infinite $N$ limit, the scramblon is a free field, and $1/N$ corrections to the OTOC arise from scramblon loop diagrams. One qualitative effect of these loops \cite{Gu:2021xaj} is to tame the exponential growth so that the OTOC saturates at a constant value (zero) after the ``scrambling time.''\footnote{See also \cite{stanford2022subleading,choi2023effective, Gao:2023wun}. In a bulk description by Einstein gravity, the scramblon is a shock wave mode of the metric, and the leading scramblon loops correspond \cite{Shenker:2014cwa} to the eikonal resummation of graviton exchange \cite{tHooft:1987vrq,Amati:1987uf,Verlinde:1991iu,Kabat:1992tb}.}

The purpose of this paper is to understand further effects of scramblon loops in the SYK model and its spatially extended versions. Our main motivation is to understand the origin of a phenomenon known as ``diffusive broadening'' from the perspective of the scramblon field theory. This effect was first observed by studying the OTOC in random circuit models~\cite{nahum2017quantum,von2018operator}, where in one spatial dimension the spatial width of the operator front was found to grow as $\sqrt{t}$.  To explore the interplay of this broadening with the large $N$ limit, \cite{Xu:2018dfp} considered a family of large-$N$ time-dependent Hamiltonians in one spatial dimension. By mapping the operator growth in these models to a noisy FKKP equation, they argued for a dramatic broadening of the operator front, with a width proportional to $\sqrt{t}/\log^{3/2}(N)$.

The model studied in~\cite{Xu:2018dfp} is very similar to a 1+1 dimensional SYK chain with time-dependent couplings (Brownian chain) and to an ordinary SYK chain at high temperature, so we expect the same behavior in these models. From the perspective of scramblon field theory, diffusive broadening implies that the exponentially growing single-scramblon contribution to the OTOC must become a bad approximation long before it becomes of order one. We find a class of loop diagrams that support this conclusion -- these diagrams describe fluctuations in which the scrambling process initially proceeds rapidly in space (via one scramblon), so that the later phase of scrambling is affected by OTOC saturation effects (with many scramblons):
\begin{align}\label{loopdiagintro}
\begin{tikzpicture}[scale=.75, rotate=0, baseline={([yshift=-0.15cm]current bounding box.center)}]
    \draw[line width=1pt, black,decorate, decoration = {snake, amplitude = 1, segment length = 6pt}] (0,0)  -- (4,4);
    \filldraw[black] (0,0) circle (0.1);% node[black, left=4] {$(0,0)$};
   \filldraw[very thick, fill = white] (4,4) circle (0.1);% node[black, left=4] {$(x,t)$};
\end{tikzpicture}
\hspace{10pt} + \hspace{10pt}
\begin{tikzpicture}[scale=.7, rotate=0, baseline={([yshift=-0.15cm]current bounding box.center)}]
    \draw[line width=1pt, black,decorate, decoration = {snake, amplitude = 1, segment length = 5pt}] (0,0) -- (2,2);
    %\draw[very thick, dashed] (2,0) -- (4,4);
    \draw[line width=1pt, black,decorate, decoration = {snake, amplitude = 1, segment length = 5pt}] (2,2) to[out = 20, in = -110] (4,4);
    \draw[line width=1pt, black,decorate, decoration = {snake, amplitude = 1, segment length = 5pt}] (2,2) to[out = 70, in = -160] (4,4);
    \filldraw[black] (0,0) circle (0.125);
    % \filldraw[red] (2,2) circle (0.125)  node[above  = 5, left = 5] {$(\xm,\tm)$};
    \filldraw[very thick, fill = white] (4,4) circle (0.125);
\end{tikzpicture}
\hspace{10pt} + \hspace{10pt}
\begin{tikzpicture}[scale=.7, rotate=0, baseline={([yshift=-0.15cm]current bounding box.center)}]
    \draw[line width=1pt, black,decorate, decoration = {snake, amplitude = 1, segment length = 5pt}] (0,0) -- (2,2);
    %\draw[very thick, dashed] (2,0) -- (4,4);
    \draw[line width=1pt, black,decorate, decoration = {snake, amplitude = 1, segment length = 5pt}] (2,2) to[out = 20, in = -110] (4,4);
    \draw[line width=1pt, black,decorate, decoration = {snake, amplitude = 1, segment length = 5pt}] (2,2) to[out = 70, in = -160] (4,4);
    \draw[line width=1pt, black,decorate, decoration = {snake, amplitude = 1, segment length = 5pt}] (2,2) to (4,4);
    \filldraw[black] (0,0) circle (0.125);
    %\filldraw[red] (2,2) circle (0.125);
    \filldraw[very thick, fill = white] (4,4) circle (0.125);
\end{tikzpicture} \hspace{10pt} + \hspace{10pt} \cdots
 \end{align}
Similar diagrams were studied in $(0+1)$d in \cite{Gu:2021xaj}. The new feature in $(1+1)$d is that for sufficiently large $x$ these diagrams require resummation before the first (tree) diagram becomes of order one. 

The Feynman-diagram method we use is less powerful than the noisy FKPP equation \cite{Xu:2018dfp}, and we are not sure how to resum enough diagrams to go beyond the breakdown time. The advantage is that the Feynman diagrams can also be used to study the SYK chain in the low-temperature regime. There, the behavior of the scramblon propagator changes \cite{gu2017local,gu2019relation} from a ``saddle point'' contribution that resembles Brownian SYK to a ``pole'' contribution that resembles graviton exchange and saturates the chaos bound \cite{Maldacena:2015waa}. Gu, Kitaev, and Zhang \cite{Gu:2021xaj} refer to the pole contribution to the scramblon propagator as coherent, in the sense that it represents a quantum amplitude for scrambling. By contrast the rest of the scramblon propagator is incoherent in the sense that it represents a probability.

We find that the pole contribution to the initial scramblon propagator cancels out in the loop diagrams (\ref{loopdiagintro}). We further find that what remains of this set of diagrams is compatible with the exponentially growing single-scramblon approximation remaining valid until an $O(1)$ distance from the front. Of course, to know whether it actually does remain valid, one would need to rule out a breakdown from some other set of diagrams, and we do not claim to have done that.\footnote{In fact, for very large separation $x \sim N$, we find another class of diagrams that will lead to a breakdown significantly before the front. We expect these are associated to a slow but nonzero broadening of order $\sqrt{t}/\sqrt{N}$.} But the results suggest to us that the the picture of scrambling as a classical stochastic process (and the associated rapid front broadening) may not be valid at low temperature.

While the spatially extended cases are the main focus of this paper, we also study the effect of scramblon loops in the $(0+1)$-dimensional Brownian SYK dot. This model has a simple collective field description that is directly the scramblon field theory -- there are no other propagating modes. We review the resummation of the leading $(\frac{1}{N} e^{\lambda_L t})^k$ effects from~\cite{Gu:2021xaj} which lead to saturation of the OTOC, and we also show how to sum a class of subleading effects proportional to $(t/N)^k$ that lead to a very mild form of diffusive broadening.

The plan of the paper is as follows. In section~\hyperref[sec:two]{two}, we study the single-scramblon exchange and loop diagrams in the (0+1)-dimensional Brownian SYK dot in the large $p$ limit. In section~\hyperref[BSYK1d]{three}, we study a $(1+1)$-dimensional chain version of the same model, and explain the breakdown of the single-scramblon exchange well before the scrambling time. In section~ \hyperref[sec:four]{four}, we study a non-Brownian version of the chain, and focus on the low-temperature regime where the ``coherent'' scrambling leads to qualitatively new effects. 

\section{Large \texorpdfstring{$p$}{p} Brownian SYK model}\label{sec:two}
The Brownian SYK model~\cite{Saad:2018bqo} is defined by the following ensemble of time-dependent Hamiltonians:
\begin{align}\label{0dBSYK}
&H(t) = \i^{p/2}\sum_{1 \le i_1<\dots < i_p \le N} J_{i_1\dots i_p}(t)\psi_{i_1}\dots \psi_{i_p}, \quad \{\psi_i,\psi_j \}=2\delta_{ij}\\
&\overline{J_{i_1\dots i_p}(t)J_{i_1\dots i_p}(t')} = \delta_{i_1i_1'}\dots\delta_{i_p i_p'}\delta(t-t')\frac{\mathcal{J}}{\lambda \binom{N}{p}}, \quad \lambda = \frac{2p^2}{N}.
\end{align}
For technical convenience, we will study the ``double scaled'' limit $p\to\infty$ and $N\to\infty$ with $\lambda$ held fixed. We then work perturbatively in $\lambda$. This is a very simple model that can be analyzed by various different methods, including ladder diagrams, chord diagrams, and the collective field description.

The collective field description can be derived by starting with the analogous description of the ordinary (not Brownian) SYK model in the double-scaled limit~\cite{Maldacena:2016hyu, Cotler:2016fpe}:
\be\label{ordinaryI}
I = \frac{1}{2\lambda}\int_0^\beta \d\tau_1\int_0^\beta \d\tau_2 \left[\frac{1}{4}\partial_1g(\tau_1,\tau_2)\partial_2g(\tau_1,\tau_2) - \mathcal{J}^2e^{g(\tau_1,\tau_2)}\right].
\ee
Here we wrote the action appropriate for computing the thermal partition function at inverse temperature $\beta$. The dynamical variable $g(\tau_1,\tau_2)$ represents  the correlations between points $\tau_1,\tau_2$, and observables can be expressed in terms of $g$. For example, an operator defined by \cite{Berkooz:2018qkz}
\be \label{wdef}
W=i^{p_W/2}\sum_{1\leq i_1 <\dots < i_{p_W}\leq N} w_{i_1\dots i_{p_W}}\psi_{i_1}\dots\psi_{i_{p_W}}, \quad \overline{w_{i_1\dots i_{p_W}}^2}=\binom{N}{p_W}^{-1}
\ee
has a two point function given by
\be
\lr W(\tau_1) W(\tau_2)\rr= \lr e^{\Delta_W g(\tau_1,\tau_2)}\rr, \quad \quad \Delta_W=\frac{p_W}{p}.
\ee
where $W(\tau) = e^{\tau H} W e^{-\tau H}$.

We will be interested in calculating OTOCs of the schematic form $\lr W(\i t)V(0)W(\i t)V(0) \rr$. To compute them, one can continue the action (\ref{ordinaryI}) to the following doubled Schwinger-Keldysh contour, e.g.
\be
\begin{tikzpicture}[scale=0.75, rotate=0, baseline={([yshift=-0.15cm]current bounding box.center)}]
  \def\xc{5}
  \def\yy{.5}
  \def\yyy{.5}
  \draw[very thick, gray, rounded corners=.5mm] (0,0) -- (\xc,0) -- (\xc, -\yy) -- (0,-1*\yy) -- (0,-2.*\yy) -- (\xc, -2.*\yy) -- (\xc, -3.0*\yy) -- (0, -3.*\yy) -- (0,-4*\yy);
  \filldraw[black] (0,0) circle (0.1) node[left] {$V(0)$};
  \filldraw[black] (0,-2*\yy) circle (0.1) node[left] {$V(\frac{\beta}{2})$};
  \draw[very thick, black, fill = white] (\xc,-1*\yy) circle (0.1) node[right] {$W(\frac{ \beta}{4}+\i t)$};
  \draw[very thick, black, fill = white] (\xc,-3*\yy) circle (0.1) node[right] {$W(\frac{ 3\beta}{4}+\i t)$};
  \draw[very thick, black] (9,.1) -- (9,-.5);
  \draw[very thick, black]  (9,-.5) -- (9.5,-.5) node[midway, above] {$\i t$};
\end{tikzpicture} 
\ee
In the Brownian case, due to lack of energy conservation, it is natural to take $\beta = 0$, so that the contour reduces to four segments that each connect time zero and $\i t$. Continuing (\ref{ordinaryI}) to this contour and then substituting $\mathcal{J}^2 \to \mathcal{J}\delta(t_1-t_2)$ to get the Brownian model, we find 
\be\label{0daction}
I = \sum_{ij}\sigma(i,j)\frac{1}{2\lambda}\int_0^t \d t_1\int_0^t\d t_2 \left[\frac{1}{4}\partial_1g_{ij}(\i t_1,\i t_2)\partial_2g_{ij}(\i t_1,\i t_2) +\delta(t_1-t_2)\mathcal{J}e^{g_{ij}(\i t_1,\i t_2)}\right].
\ee
Here $i,j$ each run over the four contours, and $\sigma(i,j)=1$ if both the $i,j$ contour go in the same direction, and equals to $-1$ otherwise. 

\subsection{Scramblon propagator}
In terms of the $g$ field on the doubled Schwinger-Keldysh contour, the OTOC is simply
\be\label{wvwvh}
\langle W(\i t)V(0)W(\i t)V(0)\rangle = \langle e^{\Delta_W g_{13}(\i t,\i t)}e^{\Delta_V g_{24}(0,0)}\rangle.
\ee
We would like to compute such correlation functions in an expansion for small $\lambda$. At leading order, one should find the saddle point of the action (\ref{0daction})
\be
\partial_1\partial_2 {\sf g}_{ij}(\i t_1,\i t_2) = 2\delta(t_1-t_2)\mathcal{J}e^{{\sf g}_{ij}(\i t_1,\i t_2)} \hspace{20pt} \implies \hspace{20pt}  {\sf g}_{ij}(\i t_1,\i t_2) = -\mathcal{J}|t_1-t_2|.
\ee
Here we use the notation ${\sf g}$ to represent the saddle point for the functional integration variable $g$. This saddle point captures the exponential decay of the two point function because it leads to $\langle W(\i t)W(0)\rangle = e^{-\Delta_W \mathcal{J}|t|}$. However, it leads to a trivial OTOC $\langle W(\i t)V(0)W(\i t)V(0)\rangle = 1$. 

To compute the OTOC at leading nontrivial order, we can expand in small fluctuations around the saddle point
\be
g_{ij}(\i t_1,\i t_2)={\sf g}_{ij}(\i t_1,\i t_2)+h_{ij}(\i t_1,\i t_2).
\ee
The action for $h$ at quadratic order is
\be
I = \sum_{i<j}\sigma(i,j)\frac{1}{\lambda}\int_0^t \d t_1\int_0^t\d t_2 \left[\frac{1}{4}\partial_1h_{ij}(\i t_1,\i t_2)\partial_2h_{ij}(\i t_1,\i t_2) + \frac{\mathcal{J}}{2}\delta(t_1-t_2)h_{ij}(\i t_1,\i t_2)^2\right].
\ee
The propagator for $h$ can be derived by solving the Green's function equation for this action, and we get an interesting perspective by solving this equation using perturbation theory in $\mathcal{J}$.\footnote{This corresponds to summing the traditional SYK ladder diagrams for this simple system.} First, if $\mathcal{J} = 0$, the action is topological and the propagator is (see appendix H of \cite{Lin:2023trc})
\be\label{freeJJ}
\langle h_{ij}(\i t_1,\i t_2) h_{i'j'}(\i t_3, \i t_4)\rangle = \begin{cases} -\lambda & \text{OTOC} \\ 0 & \text{TOC}.\end{cases}
\ee
Now, consider the term of order $\mathcal{J}$. To compute this, we expand down the term $\mathcal{J}h^2$ from the action once and connect each of the factors of $h$ to the external operators using the propagator (\ref{freeJJ}). In order to get a nonzero answer, the $\mathcal{J} h^2$ must be inserted in such a way that it is OTO relative to both the initial and final operator insertions. If the external operators are themselves in a TO configuration, there is no way to do this, so the propagator in that configuration remains zero. If the external operators are in an OTO configuration, then there are two possibilities, shown with red dots:
\be
\begin{tikzpicture}[scale=0.75, rotate=0, baseline={([yshift=-0.15cm]current bounding box.center)}]
  \def\xc{5}
  \def\yy{.5}
  \draw[very thick, gray, rounded corners=.5mm] (0,0) -- (2*\xc/3,0) -- (2*\xc/3, -\yy) -- (\xc/3,-1*\yy) -- (\xc/3,-2.*\yy) -- (\xc, -2.*\yy) -- (\xc, -3.0*\yy) -- (0, -3.*\yy);
  \filldraw[black] (0,0) circle (0.1) node[left] {$t_1$};
  \filldraw[black] (\xc/3,-2*\yy) circle (0.1) node[left] {$t_2$};
  \draw[very thick, black, fill = white] (2*\xc/3,-1*\yy) circle (0.1) node[right] {$t_3$};
  \draw[very thick, black, fill = white] (\xc,-3*\yy) circle (0.1) node[right] {$t_4$};
  \filldraw[red] (\xc/2,0) circle (0.1);
  \filldraw[red] (\xc/2,-2*\yy) circle (0.1);
  %\node[text width = 100] at (\xc/2, -4*\yy) {\begin{center}$-\lambda e^{2t} + \dots$\end{center}};
  %\draw[|-|, very thick] (\xc/3,\yy) -- (2*\xc/3,\yy) node[midway, above] {$t$};
\end{tikzpicture} \hspace{20pt} + \hspace{20pt} 
\begin{tikzpicture}[scale=0.75, rotate=0, baseline={([yshift=-0.15cm]current bounding box.center)}]
  \def\xc{5}
  \def\yy{.5}
  \draw[very thick, gray, rounded corners=.5mm] (0,0) -- (2*\xc/3,0) -- (2*\xc/3, -\yy) -- (\xc/3,-1*\yy) -- (\xc/3,-2.*\yy) -- (\xc, -2.*\yy) -- (\xc, -3.0*\yy) -- (0, -3.*\yy);
  \filldraw[black] (0,0) circle (0.1) node[left] {$t_1$};
  \filldraw[black] (\xc/3,-2*\yy) circle (0.1) node[left] {$t_2$};
  \draw[very thick, black, fill = white] (2*\xc/3,-1*\yy) circle (0.1) node[right] {$t_3$};
  \draw[very thick, black, fill = white] (\xc,-3*\yy) circle (0.1) node[right] {$t_4$};
  \filldraw[red] (\xc/2,-\yy) circle (0.1);
  \filldraw[red] (\xc/2,-3*\yy) circle (0.1);
  %\node[text width = 100] at (\xc/2, -4*\yy) {\begin{center}$-\lambda e^{2t} + \dots$\end{center}};
  %\draw[|-|, very thick] (\xc/3,\yy) -- (2*\xc/3,\yy) node[midway, above] {$t$};
\end{tikzpicture}
\ee
These dots must be between $t_2$ and $t_3$, and integrating over their location leads the $O(\mathcal{J})$ term $-\lambda(1 + 2\cdot \mathcal{J}(t_3-t_2) + \dots)$. Higher orders in the perturbation theory in $\mathcal{J}$ exponentiates and the sum is $-\lambda e^{2\mathcal{J}(t_3-t_2)}$ in the configuration sketched above.

For the calculations below, we will only need the special case involving $h_{ij}(\i t_1,\i t_2)$ with $t_1 = t_2$. To streamline the notation, we will omit one of the time arguments and the factors of $\i$ and define
\be
\h_{ij}(t) \equiv h_{ij}(\i t,\i t).
\ee
We will also set $\mathcal{J} = 1$ in the remainder of the paper, so the propagator is
\be\label{}
\langle \h_{ij}(t)\h_{i'j'}(t')\rangle = \begin{cases} -\lambda e^{2|t-t'|}  & \text{OTOC} \\ 0 & \text{TOC} \end{cases}.
\ee
The OTOC (\ref{wvwvh}) is therefore
\be\label{wvwvh2}
\langle e^{\Delta_W g_{13}(\i t,\i t)}e^{\Delta_V g_{24}(0,0)}\rangle = 1 - \Delta_W\Delta_V \lambda e^{2|t|} + O(\lambda^2).
\ee
Below, we will refer to the OTOC $\h$ propagator as a scramblon propagator and use a wavy line to represent it, e.g.
\be\label{1scramblondiag}
\begin{aligned}
 \langle \h_{13}(0) \h_{24}(t)\rangle %\hspace{10pt} = \hspace{20pt} & \begin{tikzpicture}[scale=0.45, rotate=0, baseline={([yshift=-0.15cm]current bounding box.center)}]
%   \def\xc{10}
%   \def\yy{.5}
%   \def\yyy{0.5}
%    \draw[very thick, gray!40, rounded corners=.5mm] (0,-\yy/2) -- (\xc,-\yy/2) -- (\xc, -1.5*\yy) -- (0,-1.5*\yy) -- (0,-2.5*\yy) -- (\xc, -2.5*\yy) -- (\xc, -3.5*\yy) -- (0, -3.5*\yy);
%  % \draw[very thick, gray!40, rounded corners=.5mm] (\xc,0) -- (0,0) -- (0,-\yy) -- (\xc,-\yy) -- (\xc,-\yyy - \yy) -- (0,-\yyy - \yy) -- (0,-\yyy - 2*\yy) -- (\xc,-\yyy - 2*\yy) --  (\xc,-2*\yy - 2*\yyy);
%   \draw[very thick] (0,-\yy/2) to[out = 10, in = 165] (\xc,-.5*\yyy-\yy);
%   \draw[very thick] (0,-\yyy-1.5*\yy) to[out = 10, in = 165] (\xc,-1.5*\yyy-2*\yy);
%   \filldraw[black] (0,-.5*\yy) circle (0.1);
%   \filldraw[black] (0,-1*\yyy-1.5*\yy) circle (0.1);
%   \draw[very thick, black, fill = white] (\xc,-.5*\yyy-\yy) circle (0.1);
%   \draw[very thick, black, fill = white] (\xc,-1.5*\yyy-2*\yy) circle (0.1);
%   \DrawRung{\xc/10}{-.01};
%   \DrawRung{2*\xc/10}{.1};
%   \DrawRung{3*\xc/10}{.18};
%   \DrawRung{4*\xc/10}{.18};
%   \DrawRung{5*\xc/10}{.15};
%   \DrawRung{6*\xc/10}{.06};
%   \DrawRung{7*\xc/10}{-.02};
%   \DrawRung{8*\xc/10}{-.29};
%   \DrawRung{9*\xc/10}{-.49};
% \end{tikzpicture}
 \hspace{20pt}  \longrightarrow \hspace{20pt}\begin{tikzpicture}[scale=0.6, rotate=0, baseline={([yshift=-0.15cm]current bounding box.center)}]
 \filldraw[black] (-2,0) circle (0.125) node[above] {$0$};
 \draw[line width=1pt, decorate, decoration = {snake, segment length = 7pt, amplitude = 1}] (-2,0) to (2,0);
 \draw[very thick, black, fill = white] (2,0) circle (0.125) node[above] {$t$};
\end{tikzpicture}
\end{aligned}
\ee
This notation assumes that the endpoints are in an OTOC configuration but otherwise throws away contour information, and we will have to keep track of that separately.

\eqref{wvwvh2} is the well-studied exponential growth of the OTOC at times  earlier than the scrambling time, $\lambda e^{2t} \ll 1$. In order to understand corrections to this exponential growth as we approach the scrambling time, we need to understand the loop corrections to \eqref{wvwvh2} at order $\lambda^2$ and higher. We discuss such corrections in the rest of this section.

\subsection{Loop corrections to the OTOC}

To understand the structure of the loop corrections, we will first explicitly compute all contributions to \eqref{wvwvh2} at order $\lambda^2$ in Sec. \ref{sec:loop1}. In Sec. \ref{0+1deikonal}, we will discuss how a subset of these contributions and their generalizations for higher powers of $\lambda$ give rise to the multi-scramblon resummation of \cite{stanford2022subleading,Gu:2021xaj}. We will then re-sum an additional set of contributions in Sec. \ref{0+1dcorrections}, which lead to a smearing of the multi-scramblon resummation result over a width $\sqrt{\lambda t}$. 

\subsubsection{Explicit computation of terms at order $\lambda^2$}
\label{sec:loop1}
Loop diagrams can arise either by expanding the external operators $e^{\Delta_W \h(0)}$ and $e^{\Delta_V \h(t)}$ in powers of $\h$ or by expanding down in interaction vertices from the action (\ref{0daction}), which contains the terms
\be\label{cubicInt}
I_{\text{interaction}} = \sum_{i<j}\sigma(i,j)\frac{1}{\lambda}\int_0^t \d t' \left[ \frac{1}{3!}\h_{ij}(t')^3+\frac{1}{4!}\h_{ij}(t')^4 + \dots\right].
\ee
We will start by discussing the diagrams that appear at order $\lambda^2$.

One simple diagram arises by expanding both external operators to quadratic order in $\h$ and then contracting them with free propagators:
\be
\begin{tikzpicture}[scale=0.6, rotate=0, baseline={([yshift=-0.15cm]current bounding box.center)}]
 \filldraw[black] (-2,0) circle (0.125);
 \draw[line width=1pt, decorate, decoration = {snake, segment length = 7pt, amplitude = 1}] (-2,0) to[out = 45, in = 135] (2,0);
 \draw[line width=1pt, decorate, decoration = {snake, segment length = 7pt, amplitude = 1}] (-2,0) to[out = -45, in = 215] (2,0);
 \draw[very thick, black, fill = white] (2,0) circle (0.125);
\end{tikzpicture} \hspace{20pt} = \hspace{20pt} \frac{\Delta_W^2\Delta_V^2}{2}\lambda^2 e^{4t}.
\ee

The next simplest diagrams arise by expanding one of the operators to linear order and the other to quadratic order and then using a single cubic vertex to complete the diagram:
\be\label{nextloop}
\begin{tikzpicture}[scale=0.6, rotate=0, baseline={([yshift=-0.15cm]current bounding box.center)}]
 \draw[line width=1pt, decorate, decoration = {snake, segment length = 7pt, amplitude = 1}] (-4,0) to (-2,0);
 \draw[line width=1pt, decorate, decoration = {snake, segment length = 7pt, amplitude = 1}] (-2,0) to[out = 45, in = 135] (2,0);
 \draw[line width=1pt, decorate, decoration = {snake, segment length = 7pt, amplitude = 1}] (-2,0) to[out = -45, in = 215] (2,0);
 \draw[very thick, black, fill = white] (2,0) circle (0.125);
 \filldraw[red] (-2,0) circle (0.15) node[above left] {$t'$};
 \filldraw[black] (-4,0) circle (0.125);
\end{tikzpicture} %\hspace{10pt} + \hspace{10pt} \begin{tikzpicture}[scale=0.6, rotate=0, baseline={([yshift=-0.15cm]current bounding box.center)}]
% \draw[line width=1pt, decorate, decoration = {snake, segment length = 7pt, amplitude = 1}] (2,0) to (4,0);
% \draw[line width=1pt, decorate, decoration = {snake, segment length = 7pt, amplitude = 1}] (-2,0) to[out = 45, in = 135] (2,0);
% \draw[line width=1pt, decorate, decoration = {snake, segment length = 7pt, amplitude = 1}] (-2,0) to[out = -45, in = 215] (2,0);
% \draw[very thick, black, fill = white] (4,0) circle (0.125);
% \filldraw[red] (2,0) circle (0.15) node[above right] {$t'$};
% \filldraw[black] (-2,0) circle (0.125);
%\end{tikzpicture}
\ee
Here the red dot corresponds to the cubic interaction vertex, and the black wavy lines are tree-level $h$ propagators. Because the interaction vertex (\ref{cubicInt}) involves a sum over a pair of contours $1 \le i < j \le 4$, the interaction vertex expands out to $\binom{4}{2} = 6$ terms. Most of these terms give zero, because the $\langle \h \h\rangle$ propagators vanish unless their endpoints are in an OTOC configuration. The nonzero configurations are
\be\label{contourSUMMM}
\begin{tikzpicture}[scale=0.7, rotate=0, baseline={([yshift=-0.15cm]current bounding box.center)}]
  \def\xc{5}
  \def\yy{.5}
  \draw[very thick, gray, rounded corners=.5mm] (0,0) -- (\xc,0) -- (\xc, -\yy) -- (0,-1*\yy) -- (0,-2.*\yy) -- (\xc, -2.*\yy) -- (\xc, -3.0*\yy) -- (0, -3.*\yy);
  \filldraw[black] (0,0) circle (0.1);
  \filldraw[black] (0,-1.5*\yy) circle (0.1);
  \draw[very thick, black, fill = white] (\xc,-.5*\yy) circle (0.1);
  \draw[very thick, black, fill = white] (\xc,-2.5*\yy) circle (0.1);
  \filldraw[red] (\xc/2.0,0) circle (0.1);
  \filldraw[red] (\xc/2.0,-2.0*\yy) circle (0.1);
\end{tikzpicture} \hspace{10pt} + \hspace{10pt} 
\begin{tikzpicture}[scale=0.7, rotate=0, baseline={([yshift=-0.15cm]current bounding box.center)}]
  \def\xc{5}
  \def\yy{.5}
  \draw[very thick, gray, rounded corners=.5mm] (0,0) -- (\xc,0) -- (\xc, -\yy) -- (0,-1*\yy) -- (0,-2.*\yy) -- (\xc, -2.*\yy) -- (\xc, -3.0*\yy) -- (0, -3.*\yy);
  \filldraw[black] (0,0) circle (0.1);
  \filldraw[black] (0,-1.5*\yy) circle (0.1);
  \draw[very thick, black, fill = white] (\xc,-.5*\yy) circle (0.1);
  \draw[very thick, black, fill = white] (\xc,-2.5*\yy) circle (0.1);
  \filldraw[red] (\xc/2.0,-\yy) circle (0.1);
  \filldraw[red] (\xc/2.0,-3.0*\yy) circle (0.1);
\end{tikzpicture}
\ee
In this diagram we are showing the OTOC timefold. The red dots indicate the two contours $i,j$ that appear in the interaction at time $t'$. We have omitted the propagators for clarity, showing only the locations of the interactions. The sum over the two diagrams gives a factor of two, and we end up with
\begin{align}
\begin{tikzpicture}[scale=0.6, rotate=0, baseline={([yshift=-0.15cm]current bounding box.center)}]
 \draw[line width=1pt, decorate, decoration = {snake, segment length = 7pt, amplitude = 1}] (-4,0) to (-2,0);
 \draw[line width=1pt, decorate, decoration = {snake, segment length = 7pt, amplitude = 1}] (-2,0) to[out = 45, in = 135] (2,0);
 \draw[line width=1pt, decorate, decoration = {snake, segment length = 7pt, amplitude = 1}] (-2,0) to[out = -45, in = 215] (2,0);
 \draw[very thick, black, fill = white] (2,0) circle (0.125);
 \filldraw[red] (-2,0) circle (0.15) node[above left] {$t'$};
 \filldraw[black] (-4,0) circle (0.125);
\end{tikzpicture} \hspace{10pt} &=\hspace{6pt} -\frac{2}{\lambda}\cdot\frac{\Delta_W\Delta_V^2}{2}\int_0^t \d t' \frac{3\cdot 2}{3!} \langle \h(0)\h(t')\rangle \langle \h(t')\h(t)\rangle^2\\
&=\hspace{6pt} \lambda^2\Delta_W\Delta_V^2\int_0^t \d t' e^{2t' + 4(t-t')}\\
&=\hspace{6pt} \lambda^2\Delta_W\Delta_V^2\frac{e^{4t} - e^{2t}}{2}.
\end{align}

Finally, we have the diagram
\be\label{onelooppropagator}
\begin{tikzpicture}[scale=.65, rotate=0, baseline={([yshift=-0.15cm]current bounding box.center)}]
    \draw[line width=1pt, black, decorate, decoration = {snake, amplitude = 1, segment length = 7pt}] (0,0) -- (3,0);
    \draw[line width=1pt, black, decorate, decoration = {snake, amplitude = 1, segment length = 7pt}] (6,0) arc (0:360:1.5);
    [very thick, black, decorate, decoration = {snake, amplitude = 1, segment length = 7pt}] (0,0) -- (3,0);
     \draw[line width=1pt, black, decorate, decoration = {snake, amplitude = 1, segment length = 7pt}] (6,0) -- (9,0);
     \filldraw[black] (0,0) circle (.125);
     \filldraw[red] (3,0) circle (.125) node[above left] {$t'$};
    \filldraw[blue] (6,0) circle (.125) node[above right] {$t''$};
     \draw[black, very thick, fill = white] (9,0) circle (.125);
\end{tikzpicture}
\ee
In principle the sum over contours contains $6\cdot 6 = 36$ terms in total. Many of these terms give zero, because the $\langle \h \h\rangle$ propagators vanish unless their endpoints are in an OTOC configuration. In fact, if $t'' > t'$ then the entire contribution can be reduced to the following four diagrams, corresponding to two independent copies of the sum in (\ref{contourSUMMM}):
\be
\begin{tikzpicture}[scale=0.65, rotate=0, baseline={([yshift=-0.15cm]current bounding box.center)}]
  \def\xc{5}
  \def\yy{.5}
  \draw[very thick, gray, rounded corners=.5mm] (0,0) -- (\xc,0) -- (\xc, -\yy) -- (0,-1*\yy) -- (0,-2.*\yy) -- (\xc, -2.*\yy) -- (\xc, -3.0*\yy) -- (0, -3.*\yy);
  \filldraw[black] (0,0) circle (0.1);
  \filldraw[black] (0,-1.5*\yy) circle (0.1);
  \draw[very thick, black, fill = white] (\xc,-.5*\yy) circle (0.1);
  \draw[very thick, black, fill = white] (\xc,-2.5*\yy) circle (0.1);
  \filldraw[red] (\xc/3.0,0) circle (0.1);
  \filldraw[red] (\xc/3.0,-2.0*\yy) circle (0.1);
  \filldraw[blue] (2*\xc/3.0,-0*\yy) circle (0.1);
  \filldraw[blue] (2*\xc/3.0,-2*\yy) circle (0.1);
\end{tikzpicture} \hspace{10pt} + \hspace{10pt} 
\begin{tikzpicture}[scale=0.65, rotate=0, baseline={([yshift=-0.15cm]current bounding box.center)}]
  \def\xc{5}
  \def\yy{.5}
  \draw[very thick, gray, rounded corners=.5mm] (0,0) -- (\xc,0) -- (\xc, -\yy) -- (0,-1*\yy) -- (0,-2.*\yy) -- (\xc, -2.*\yy) -- (\xc, -3.0*\yy) -- (0, -3.*\yy);
  \filldraw[black] (0,0) circle (0.1);
  \filldraw[black] (0,-1.5*\yy) circle (0.1);
  \draw[very thick, black, fill = white] (\xc,-.5*\yy) circle (0.1);
  \draw[very thick, black, fill = white] (\xc,-2.5*\yy) circle (0.1);
  \filldraw[red] (\xc/3.0,0) circle (0.1);
  \filldraw[red] (\xc/3.0,-2.0*\yy) circle (0.1);
  \filldraw[blue] (2*\xc/3.0,-1*\yy) circle (0.1);
  \filldraw[blue] (2*\xc/3.0,-3*\yy) circle (0.1);
\end{tikzpicture} \hspace{10pt} + \hspace{10pt}
\begin{tikzpicture}[scale=0.65, rotate=0, baseline={([yshift=-0.15cm]current bounding box.center)}]
  \def\xc{5}
  \def\yy{.5}
  \draw[very thick, gray, rounded corners=.5mm] (0,0) -- (\xc,0) -- (\xc, -\yy) -- (0,-1*\yy) -- (0,-2.*\yy) -- (\xc, -2.*\yy) -- (\xc, -3.0*\yy) -- (0, -3.*\yy);
  \filldraw[black] (0,0) circle (0.1);
  \filldraw[black] (0,-1.5*\yy) circle (0.1);
  \draw[very thick, black, fill = white] (\xc,-.5*\yy) circle (0.1);
  \draw[very thick, black, fill = white] (\xc,-2.5*\yy) circle (0.1);
  \filldraw[red] (\xc/3.0,-1*\yy) circle (0.1);
  \filldraw[red] (\xc/3.0,-3.0*\yy) circle (0.1);
  \filldraw[blue] (2*\xc/3.0,-0*\yy) circle (0.1);
  \filldraw[blue] (2*\xc/3.0,-2*\yy) circle (0.1);
\end{tikzpicture} \hspace{10pt} + \hspace{10pt} 
\begin{tikzpicture}[scale=0.65, rotate=0, baseline={([yshift=-0.15cm]current bounding box.center)}]
  \def\xc{5}
  \def\yy{.5}
  \draw[very thick, gray, rounded corners=.5mm] (0,0) -- (\xc,0) -- (\xc, -\yy) -- (0,-1*\yy) -- (0,-2.*\yy) -- (\xc, -2.*\yy) -- (\xc, -3.0*\yy) -- (0, -3.*\yy);
  \filldraw[black] (0,0) circle (0.1);
  \filldraw[black] (0,-1.5*\yy) circle (0.1);
  \draw[very thick, black, fill = white] (\xc,-.5*\yy) circle (0.1);
  \draw[very thick, black, fill = white] (\xc,-2.5*\yy) circle (0.1);
  \filldraw[red] (\xc/3.0,-1*\yy) circle (0.1);
  \filldraw[red] (\xc/3.0,-3.0*\yy) circle (0.1);
  \filldraw[blue] (2*\xc/3.0,-1*\yy) circle (0.1);
  \filldraw[blue] (2*\xc/3.0,-3*\yy) circle (0.1);
\end{tikzpicture}\notag\vspace{8pt}
\ee
If $t'' < t'$, there will also be some cancellations between nonzero diagrams, for example
\be
\begin{tikzpicture}[scale=0.7, rotate=0, baseline={([yshift=-0.15cm]current bounding box.center)}]
  \def\xc{5}
  \def\yy{.5}
  \def\yyy{.5}
  \draw[very thick, gray, rounded corners=.5mm] (0,0) -- (\xc,0) -- (\xc, -\yy) -- (0,-1*\yy) -- (0,-2.*\yy) -- (\xc, -2.*\yy) -- (\xc, -3.0*\yy) -- (0, -3.*\yy);
  \filldraw[black] (0,0) circle (0.1);
  \filldraw[black] (0,-1.5*\yy) circle (0.1);
  \draw[very thick, black, fill = white] (\xc,-.5*\yy) circle (0.1);
  \draw[very thick, black, fill = white] (\xc,-2.5*\yy) circle (0.1);
  \filldraw[red] (2*\xc/3.0,0) circle (0.1);
  \filldraw[red] (2*\xc/3.0,-3.0*\yy) circle (0.1);
  \filldraw[blue] (\xc/3.0,-\yy) circle (0.1);
  \filldraw[blue] (\xc/3.0,-3*\yy) circle (0.1);
\end{tikzpicture} \hspace{20pt} + \hspace{20pt} 
\begin{tikzpicture}[scale=0.7, rotate=0, baseline={([yshift=-0.15cm]current bounding box.center)}]
  \def\xc{5}
  \def\yy{.5}
  \draw[very thick, gray, rounded corners=.5mm] (0,0) -- (\xc,0) -- (\xc, -\yy) -- (0,-1*\yy) -- (0,-2.*\yy) -- (\xc, -2.*\yy) -- (\xc, -3.0*\yy) -- (0, -3.*\yy);
  \filldraw[black] (0,0) circle (0.1);
  \filldraw[black] (0,-1.5*\yy) circle (0.1);
  \draw[very thick, black, fill = white] (\xc,-.5*\yy) circle (0.1);
  \draw[very thick, black, fill = white] (\xc,-2.5*\yy) circle (0.1);
  \filldraw[red] (2*\xc/3.0,0) circle (0.1);
  \filldraw[red] (2*\xc/3.0,-3.0*\yy) circle (0.1);
  \filldraw[blue] (\xc/3.0,-2.0*\yy) circle (0.1);
  \filldraw[blue] (\xc/3.0,-3*\yy) circle (0.1);
\end{tikzpicture} \hspace{20pt} = \hspace{20pt} 0. \vspace{8pt}
\ee
The only difference between the two diagrams is the location of one of the blue interaction points. The contour ordering of this blue point remains the same relative to the red and white points.~\footnote{In the diagram (\ref{onelooppropagator}) the blue vertex does not connect to the black point directly, so the change in contour ordering between the black and blue points does not make a difference.} This means that the two diagrams are equal except for a relative minus sign due to the change of sign of $\sigma(i,j)$, and they therefore cancel. In fact, one finds that if $t'' < t'$ then all diagrams cancel. 

Adding together the four equal cases for $t'' > t'$, the complete one-loop correction is:
\begin{align}\label{singlescramblonloop}
\begin{tikzpicture}[scale=.65, rotate=0, baseline={([yshift=-0.15cm]current bounding box.center)}]
    \draw[line width=1pt, black, decorate, decoration = {snake, amplitude = 1, segment length = 7pt}] (0,0) -- (3,0);
    \draw[line width=1pt, black, decorate, decoration = {snake, amplitude = 1, segment length = 7pt}] (6,0) arc (0:360:1.5);
    [very thick, black, decorate, decoration = {snake, amplitude = 1, segment length = 7pt}] (0,0) -- (3,0);
     \draw[line width=1pt, black, decorate, decoration = {snake, amplitude = 1, segment length = 7pt}] (6,0) -- (9,0);
     \filldraw[black] (0,0) circle (.125);
     \filldraw[red] (3,0) circle (.125) node[above left] {$t'$};
    \filldraw[blue] (6,0) circle (.125) node[above right] {$t''$};
     \draw[black, very thick, fill = white] (9,0) circle (.125);
\end{tikzpicture} \hspace{10pt} %= \hspace{10pt} 4\cdot \frac{1}{\lambda^2}\int_0^t \d t'' \int_0^{t''}\d t'\frac{3\cdot 3\cdot 2}{(3!)^2} \langle h(0,0)h(t',t')\rangle \langle h(t',t')h(t'',t'')\rangle^2\langle h(t'',t'')h(t,t)\rangle\notag\\
%&=4\cdot \frac{\lambda^2}{2}\int_0^{t}\d t'' \int_0^{t''}\d t' e^{2t'}e^{4(t''-t')}e^{2(t-t'')}\\
&= \hspace{6pt} \lambda^2\Delta_V\Delta_W \frac{e^{4t} - 2te^{2t} - e^{2t}}{2}.
\end{align}
The leading part of this result, proportional to $\lambda^2 e^{4t}$, arises from a part of the integration space where the interaction vertices are close to the endpoints:
\be
\begin{tikzpicture}[scale=.65, rotate=0, baseline={([yshift=-0.15cm]current bounding box.center)}]
    \draw[line width=1pt, black, decorate, decoration = {snake, amplitude = 1, segment length = 7pt}] (0,0) -- (1,0);
    \draw[line width=1pt, black, decorate, decoration = {snake, amplitude = 1, segment length = 7pt}] (4.5,0) ellipse (3.5cm and 1cm);
     \draw[line width=1pt, black, decorate, decoration = {snake, amplitude = 1, segment length = 7pt}] (8,0) -- (9,0);
     \filldraw[black] (0,0) circle (.125);
     \filldraw[red] (1,0) circle (.125);
    \filldraw[blue] (8,0) circle (.125) ;
     \draw[black, very thick, fill = white] (9,0) circle (.125);
\end{tikzpicture}
\ee
The first subleading contribution, proportional to $-\lambda^2 t e^{2t}$, results from the part of the integral where the loop is small, and the factor of $t$ is due to a zero mode that translates the entire loop.
\be\label{smallLoop}
\begin{tikzpicture}[scale=.65, rotate=0, baseline={([yshift=-0.15cm]current bounding box.center)}]
    \draw[line width=1pt, black, decorate, decoration = {snake, amplitude = 1, segment length = 7pt}] (0,0) -- (2.5,0);
    \draw[line width=1pt, black, decorate, decoration = {snake, amplitude = 1, segment length = 5.5pt}] (3,0) ellipse (.5cm and .5cm);
     \draw[line width=1pt, black, decorate, decoration = {snake, amplitude = 1, segment length = 7pt}] (3.5,0) -- (9,0);
     \filldraw[black] (0,0) circle (.125);
     \filldraw[red] (2.5,0) circle (.125);
    \filldraw[blue] (3.5,0) circle (.125) ;
     \draw[black, very thick, fill = white] (9,0) circle (.125);
\end{tikzpicture}
\ee

\subsubsection{Multi-scramblon resummation}\label{0+1deikonal}
The tree-level single-scramblon exchange is of order $\lambda e^{2t}$. At order $\lambda^2$, we have just seen that there are contributions of order $(\lambda e^{2t})^2$ coming from several diagrams. In general, at order $\lambda^k$ the leading terms are of order $(\lambda e^{2t})^k$, and the formula that resums them is known \cite{stanford2022subleading,Gu:2021xaj}:
\be\label{eikonalresum}
\begin{aligned}1 \hspace{5pt} + \hspace{5pt}
\begin{tikzpicture}[scale=.35, rotate=0, baseline={([yshift=-0.15cm]current bounding box.center)}]
    \draw[line width=1pt, black,decorate, decoration = {snake, amplitude = 1, segment length = 7pt}] (0,0) -- (0,4);
    \filldraw[black] (0,0) circle (0.125); % node[black, left=4] {$0$};
    \draw[very thick, black, fill = white] (0,4) circle (0.125);% node[black, left=4] {$t$};
    \end{tikzpicture}
\hspace{5pt} + \hspace{5pt}
\begin{tikzpicture}[scale=.35, rotate=0, baseline={([yshift=-0.15cm]current bounding box.center)}]
 %   \draw[line width=1pt, black,decorate, decoration = {snake, amplitude = 1, segment length = 6pt}] (0,0) -- (3,1);
    %\draw[very thick, dashed] (2,0) -- (4,4);
    \draw[line width=1pt, black,decorate, decoration = {snake, amplitude = 1, segment length = 6pt}] (0,0) to[out = 45, in = -45] (0,4);
    \draw[line width=1pt, black,decorate, decoration = {snake, amplitude = 1, segment length = 6pt}] (0,0) to[out = 140, in = -140] (0,4);
%    \filldraw[red]  (3,1) circle (0.15);
    \filldraw[black] (0,0) circle (0.125);
    \draw[very thick, black, fill = white] (0,4) circle (0.125);
%    \DrawBracket{1.5}{.5}{18.43}
 \draw[very thick, fill=gray!20] (0,4) ellipse (1 and 0.5);
      \draw[very thick, fill=gray!100] (0,0) ellipse (1 and 0.5);
\end{tikzpicture}
\hspace{5pt} + \hspace{5pt}
\begin{tikzpicture}[scale=.35, rotate=0, baseline={([yshift=-0.15cm]current bounding box.center)}]
%    \draw[line width=1pt, black,decorate, decoration = {snake, amplitude = 1, segment length = 6pt}] (0,0) -- (3,1);
    %\draw[very thick, dashed] (2,0) -- (4,4);
    \draw[line width=1pt, black,decorate, decoration = {snake, amplitude = 1, segment length = 6pt}] (0,0) to[out = 45, in = -45] (0,4);
    \draw[line width=1pt, black,decorate, decoration = {snake, amplitude = 1, segment length = 6pt}] (0,0) to[out = 140, in = -140] (0,4);
    \draw[line width=1pt, black,decorate, decoration = {snake, amplitude = 1, segment length = 6pt}] (0,0) to (0,4);
    \filldraw[black] (0,0) circle (0.125);
    \draw[very thick, black, fill = white] (0,4) circle (0.125);
%    \DrawBracket{1.5}{.5}{18.43}
 \draw[very thick, fill=gray!20] (0,4) ellipse (1 and 0.5);
      \draw[very thick, fill=gray!100] (0,0) ellipse (1 and 0.5);
\end{tikzpicture} \hspace{5pt} + \hspace{5pt} \cdots \hspace{10pt}&= \hspace{10pt}{}_2F_0(\Delta_W,\Delta_V,-\lambda e^{2|t|}) \\
& =  \sum_{k=0}^{\infty} \frac{(\Delta_W)_k (\Delta_{V})_k}{k!} \, (-\lambda e^{2|t|})^k  
 \end{aligned}
 \ee
 where $(x)_n = x (x+1)...(x+n-1)$. On the LHS of (\ref{eikonalresum}) we are using the ``blob'' notation from \cite{Gu:2021xaj}. The leading contributions $(\lambda e^{2t})^k$ arise from regions in loop integration space in which the loops are almost as large as possible, so all interaction vertices are confined near the initial or final operator insertions. In between we have $k$ tree-level scramblons propagating. The interactions can be absorbed into slightly nonlocal-in-time vertices (blobs) that couple these scramblons to the external operators. These vertices can be computed by studying the response of a two point function to a small classical perturbation \cite{Gu:2021xaj}. In this sense, these diagrams do not correspond to genuine interactions of scramblons, but instead to free scramblons coupled nonlinearly to the external sources.

 The ${}_2F_0$ function in (\ref{eikonalresum}) is an asymptotic series and its correct resummation is the function $(\lambda e^{2|t|})^{-\Delta_W}U(\Delta_W,1+\Delta_W-\Delta_V,\lambda^{-1} e^{-2|t|})$ where $U$ is the confluent hypergeometric function. This converges to zero for late time $\lambda e^{2t} \gg 1$ -- the qualitative effect of the leading $(\lambda e^{2t})^k$ effects is to cut off the exponential growth of the single-scramblon contribution and to make the full OTOC saturate at zero.

\subsubsection{Corrections proportional to powers of $\lambda t$}\label{0+1dcorrections}
So far we have seen how to compute the OTOC including all powers of $(\lambda e^{2t})^k$. There is an interesting class of corrections to this, proportional to further powers $(\lambda t)^m$, and we would like to show how to resum these corrections. The idea is that if we start with the ``blob'' diagrams in (\ref{eikonalresum}), then in order to get a correction proportional to $t$, we need to have a loop diagram that can attach itself anywhere along the extended portion of the diagrams, where there are $k$ scramblons propagating between the blobs.

One possibility is to replace any of the $k$ scramblon propagators with a small loop (\ref{smallLoop}) which represents the linear in $t$ part of (\ref{singlescramblonloop}). By itself, this would correspond to a correction
\be
e^{2kt} \to e^{2kt}\cdot (1 + k \lambda t + \dots).
\ee
in \eqref{eikonalresum}. However, we also need to account for the possibility of a linear in $t$ contribution arising from a loop diagram that affects more than one scramblon propagator. To determine this, let's consider the loop corrections to a pair of scramblon propagators. One possibility is to simply decorate one of the propagators by a one-loop correction:
\begin{align}
\begin{tikzpicture}[scale=.65, rotate=0, baseline={([yshift=-0.15cm]current bounding box.center)}]
    \draw[line width=1pt, black, decorate, decoration = {snake, amplitude = 1, segment length = 7pt}] (0,0) -- (2,0);
    \draw[line width=1pt, black, decorate, decoration = {snake, amplitude = 1, segment length = 7pt}] (3,0) ellipse (1cm and 1cm);
     \draw[line width=1pt, black, decorate, decoration = {snake, amplitude = 1, segment length = 7pt}] (4,0) -- (6,0);
    \draw[line width=1pt, black, decorate, decoration = {snake, amplitude = 1, segment length = 7pt}] (0,-1.5) -- (6,-1.5);
    %\filldraw[black] (0,0) circle (.125);
    %\filldraw[black] (0,-1.5) circle (.125);
   % \filldraw[red] (2,0) circle (.125);
   % \filldraw[blue] (4,0) circle (.125) ;
   % \draw[black, very thick, fill = white] (6,0) circle (.125);
   % \draw[black, very thick, fill = white] (6,-1.5) circle (.125);
\end{tikzpicture} \hspace{10pt} &= \hspace{10pt} 2\cdot 2\cdot \langle h(0,0)h(t,t)\rangle_{\text{1 loop}}\cdot\langle h(0,0)h(t,t)\rangle\\
&= \hspace{10pt} \lambda^3(-2 e^{6t} + 4 t e^{4t} +2 e^{4t}). \label{232}
\end{align}
Here one factor of two is because the loop could happen on either propagator, and the other factor of two is because the two propagators could cross.

A more interesting possibility is an ``H'' diagram 
\begin{align}\label{0+1dHdiag}
\begin{tikzpicture}[scale=.65, rotate=0, baseline={([yshift=-0.15cm]current bounding box.center)}]
    \draw[line width=1pt, black, decorate, decoration = {snake, amplitude = 1, segment length = 7pt}] (0,0) -- (5,0);
    \draw[line width=1pt, black, decorate, decoration = {snake, amplitude = 1, segment length = 7pt}] (2,0) -- (3,-1.5);
    \draw[line width=1pt, black, decorate, decoration = {snake, amplitude = 1, segment length = 7pt}] (0,-1.5) -- (5,-1.5);
    %\filldraw[black] (0,0) circle (.125);
    %\filldraw[black] (0,-1.5) circle (.125);
   % \filldraw[red] (2,0) circle (.125);
   % \filldraw[blue] (4,0) circle (.125) ;
   % \draw[black, very thick, fill = white] (6,0) circle (.125);
   % \draw[black, very thick, fill = white] (6,-1.5) circle (.125);
\end{tikzpicture} \hspace{10pt} &= \hspace{10pt} 2^3\cdot (3!)^2\cdot \frac{-\lambda^3 e^{4t}}{(3!)^2}\int_0^t \d t'' \int_0^t \d t' e^{2|t''-t'|}\\
&= \hspace{10pt} \lambda^3(-4e^{6t} + 8t e^{4t} + 4 e^{4t}).
\end{align}
One factor of two is due to the possibility of a crossed diagram. The other two factors of two arise from the sum over contours, one factor of two for each interaction vertex (as in the one-loop correction to the single propagator).

A third possibility is
\begin{align}\label{0+1dtdiag}
\begin{tikzpicture}[scale=.65, rotate=0, baseline={([yshift=-0.15cm]current bounding box.center)}]
    \draw[line width=1pt, black, decorate, decoration = {snake, amplitude = 1, segment length = 7pt}] (0,0) -- (.75,-.75) -- (0,-1.5);
    \draw[line width=1pt, black, decorate, decoration = {snake, amplitude = 1, segment length = 7pt}] (.75,-.75) -- (4-.75,-.75);
    \draw[line width=1pt, black, decorate, decoration = {snake, amplitude = 1, segment length = 7pt}] (4,0) -- (4-.75,-.75) -- (4, -1.5);
    %\filldraw[black] (0,0) circle (.125);
    %\filldraw[black] (0,-1.5) circle (.125);
   % \filldraw[red] (2,0) circle (.125);
   % \filldraw[blue] (4,0) circle (.125) ;
   % \draw[black, very thick, fill = white] (6,0) circle (.125);
   % \draw[black, very thick, fill = white] (6,-1.5) circle (.125);
\end{tikzpicture} \hspace{10pt} &= \hspace{10pt} 2^2\cdot (3!)^2\cdot \frac{-\lambda^3 e^{4t}}{(3!)^2}\int_0^t \d t'' \int_0^{t''} \d t' e^{-2(t''-t')}\\
&= \hspace{10pt} \lambda^3(-2te^{4t} +e^{4t} - e^{2t}).
\end{align}
Here the two factors of two arise from the sum over contours for the two interaction vertices. Finally, we have a diagram that uses a quartic interaction vertex
\begin{align}\label{0+1dcross}
\begin{tikzpicture}[scale=.65, rotate=0, baseline={([yshift=-0.15cm]current bounding box.center)}]
    \draw[line width=1pt, black, decorate, decoration = {snake, amplitude = 1, segment length = 7pt}] (0,0) -- (3,-1.5);
    \draw[line width=1pt, black, decorate, decoration = {snake, amplitude = 1, segment length = 7pt}] (0,-1.5) -- (3, 0);
    %\filldraw[black] (0,0) circle (.125);
    %\filldraw[black] (0,-1.5) circle (.125);
   % \filldraw[red] (2,0) circle (.125);
   % \filldraw[blue] (4,0) circle (.125) ;
   % \draw[black, very thick, fill = white] (6,0) circle (.125);
   % \draw[black, very thick, fill = white] (6,-1.5) circle (.125);
\end{tikzpicture} \hspace{10pt} &= \hspace{10pt} 2\cdot 4!\cdot \frac{-\lambda^3 e^{4t}}{4!}\int_0^t \d t'\\
&= \hspace{10pt} -2\lambda^3te^{4t}. \label{238}
\end{align}
Where, again, the factor of two is due to the sum over contours for the interaction vertex. Adding these up and including the tree-level exchange, we find
\be
\langle \h(0)^2 \h(t)^2\rangle = 2\lambda^2 e^{4t} + \lambda^3(-6 e^{6t} + 8 t e^{4t} + 7 e^{4t} - e^{2t}) + O(\lambda^4).
\ee
The tree-level term $\lambda^2 e^{4t}$ and the term of order $\lambda^3 e^{6t}$ will both be captured by (\ref{eikonalresum}), and in particular the $\lambda^3 e^{6t}$ piece is part of the three-scramblon contribution. The interesting term for our current purposes is the term $8\lambda^3 t e^{4t}$. This is the leading loop correction to the two-scramblon contribution.

Now, consider a diagram with $k$ scramblons. A general one-loop correction will consist of a correction to any one of the propagators or a one-loop correction to any pair. So we expect the correction to be of the form
\be
e^{2kt} \to e^{2kt} \left\{1 + \lambda t\left[\# k + \#' \binom{k}{2}\right] + O(\lambda)O(t^0) + O(\lambda^2)\right\}.
\ee
By matching to the answer $-\lambda e^{2t}\to -\lambda e^{2t}(1+\lambda t)$ for a single scramblon and $2\lambda^2e^{4t}\to 2\lambda^2e^{4t}(1 + 4\lambda t)$ for two scramblons, we conclude $\# = 1$, $\#' = 2$ so the correction for $k$ scramblons will be
\be
e^{2kt} \to e^{2kt} \left\{1 + k^2\lambda t + O(\lambda)O(t^0) + O(\lambda^2)\right\}.
\ee

So far, we have considered a single loop. More generally, we can get corrections to the term involving $(\lambda e^{2t})^k$ in \eqref{eikonalresum} that are proportional to $(\lambda t)^m$ by adding a total of $m$ loops. These can either be loop corrections to single propagators or one-loop corrections to pairs. We saw that the different options for a one-loop correction to $k$ scramblons together give a factor of $k^2\lambda t$, so the case with $m$ such corrections gives $\frac{(k^2\lambda t)^m}{m!}$, where the $1/m!$ comes from fixing the order of times in the loop integral to avoid overcounting. Then summing over all $m$, we get the following correction to the factor of $e^{2kt}$: 
\be 
e^{2kt} \rightarrow e^{2kt} \sum_{m=0}^{\infty} \frac{(k^2 \lambda t)^m}{m!}  = e^{2kt} e^{\lambda t k^2} \,. 
\ee

Now putting these corrections into the asymptotic series obtained from the multi-scramblon resummation, we get a result 
\begin{align} 
\braket{e^{\Delta_W g_{13}(\i t, \i t)} e^{\Delta_W g_{13}(0, 0)} }  &= \sum_{k=0}^{\infty} V_k \left(\lambda e^{2t}\right)^k e^{\lambda t k^2}  \label{exp}\\ 
& =  \int_{-\infty}^\infty \frac{\d t'}{\sqrt{\pi \lambda t}} e^{2k(t'-t) - \frac{(t-t')^2}{\lambda t}}  \left( \sum_{k=0}^{\infty} V_k \left(\lambda e^{2t}\right)^k  \right)  \label{smearing}
\end{align} 
where 
\be 
V_k = \frac{(\Delta_W)_k (\Delta_{V})_k}{k!} + O(\lambda) \, . 
\ee
In principle, if we keep terms up to order $(\lambda t)^m$ for $m \geq 2$ in the expansion of the exponential $e^{\lambda t k^2}$, then for consistency we should also include terms of order $O(\lambda)$ and higher in $V_k$. An example of a higher-order correction to $V_k$ for $k = 1$ is the last term of \eqref{singlescramblonloop}. 

Eq.~\eqref{smearing} shows that the effect of the loop corrections discussed here is to smear the result from the leading multi-scramblon resummation (\ref{eikonalresum}) over a width $\sqrt{\lambda t}$. This can be considered a mild version of the front-broadening effect. It is mild because at the scrambling time $t\sim \log (1/\lambda)$, the degree of broadening is of order $\sqrt{\lambda \log (1/\lambda)}$ which is small for small $\lambda$.

\section{Large \texorpdfstring{$p$}{p} Brownian SYK chain}\label{BSYK1d}
We now consider a system with spatial locality, using a modified version of the model proposed in \cite{gu2017local}. The Hamiltonian has two types of terms. One couples $p$ fermions within a site $x$, the other couples $p/2$ fermions to $p/2$ fermions in adjacent sites.
\be
\includegraphics[width = .55\textwidth]{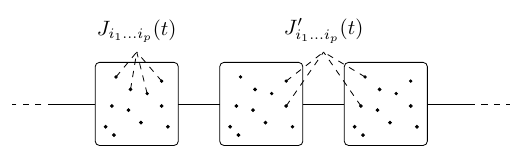} 
\ee

The full Hamiltonian is
\begin{align}
H(t) &= \i^{p/2}\sum_x\sum_{{\tiny1 \le i_1<\dots < i_p \le N}} J_{i_1\dots i_p,x}(t)\psi_{i_1,x}\dots \psi_{i_p,x}\\&+ \i^{p/2}\sum_x\sum_{\tiny{\begin{aligned}&1 \le i_1<\dots < i_{p/2} \le N\\&1 \le i_{p/2+1}<\dots < i_{p} \le N\end{aligned}}} J'_{i_1\dots i_p,x}(t)\psi_{i_1,x}\dots\psi_{i_{p/2},x} \psi_{i_{p/2+1},x+1} \psi_{i_p,x+1}\end{align}
where the couplings satisfy
\begin{align}
&\overline{J_{i_1\dots i_p,x}(t)J_{i_1\dots i_p,x'}(t')} = \delta_{x,x'}\delta_{i_1i_1'}\dots\delta_{i_p i_p'}\delta(t-t')\frac{(1-a)\mathcal{J}}{\lambda \binom{N}{p}}, \quad \\&\overline{J'_{i_1\dots i_p,x}(t)J'_{i_1\dots i_p,x'}(t')}=\delta_{x,x'}\delta_{i_1i_1'}\dots\delta_{i_p i_p'}\delta(t-t')\frac{a\mathcal{J}}{\lambda \binom{N}{p}}
\end{align}
for some $0\leq a \leq 1$. As in the previous section, let us set $\mathcal{J}=1$. 
The collective field action for this model, generalizing (\ref{0daction}) is
\be\label{1daction}
\begin{aligned}
I = \sum_{ij}\sigma(i,j)\sum_x \frac{1}{2\lambda}\int_0^t \d t_1\int_0^t\d t_2 \Bigg[&\frac{1}{4}\partial_1g_{ij}(\i t_1,\i t_2;x)\partial_2g_{ij}(\i t_1,\i t_2;x) \\ &\hspace{-80pt}+ \delta(t_1-t_2)\left((1-a)e^{g_{ij}(\i t_1,\i t_2;x)} + a e^{\frac{g_{ij}(\i t_1,\i t_2;x)+g_{ij}(\i t_1,\i t_2;x+1)}{2}}\right)\Bigg].
\end{aligned}
\ee
Note that the field $g_{ij}(t_1,t_2;x)$ is bilocal in time but local in space. The interaction term is local in both space and time because of the Brownian nature of the couplings.

\subsection{Scramblon propagator}
The saddle point is the familiar $(0+1)$d saddle point for each value of $x$:
\be
g_{*,ij}(\i t_1,\i t_2;x) = -|t_1-t_2|.
\ee
We can then expand $g_{ij}(\i t_1,\i t_2;x) = -|t_1-t_2| + h_{ij}(\i t_1,\i t_2;x)$ and write the quadratic action for $h$. This is simplest to express in momentum space 
\be
h(\i t_1,\i t_2;p) = \sum_x e^{-\i p x} h(\i t_1,\i t_2;x), \hspace{20pt} h(\i t_1,\i t_2;x) = \int_{-\pi}^\pi \frac{\d p}{2\pi}e^{\i p x} h(\i t_1,\i t_2;p).
\ee
One finds
\be
I_{\rm quad}=\sum_{ij}\frac{\sigma(i,j)}{2\lambda}\int_{-\pi}^{\pi}\frac{\d p}{2\pi}\int\hspace{-8pt}\int_0^t \d t_1\d t_2 h_{ij}(t_1,t_2;p)\left(-\frac{\partial_1\partial_2}{4} + \frac{\delta(t_1-t_2)}{4}\lambda_L(p)\right)h_{ij}(t_1,t_2;-p),
\ee
where
\be\label{lambdaL}
\lambda_L(p)=2\left(1-a\sin^2\frac{p}{2}\right).
\ee
For each value of $p$ this is the same as the $(0+1)$d case, with a rescaling of time. This implies that (using the notation $\h(x,t) \equiv h(\i t,\i t;x)$ as before, and dropping the contour indices for clarity)
\be\label{prop_p}
\langle \h(p,t)\h(-p,t')\rangle = \begin{cases} -\lambda e^{\lambda_{L}(p)|t-t'|}  & \text{OTOC} \\ 0 & \text{TOC} \end{cases}.
\ee
The propagator in the space domain is obtained by a Fourier transform
\be\label{1d1scramblon}
\langle \h(x,t)\h(0,0)\rangle = -\lambda\int_{-\pi}^\pi \frac{\d p}{2\pi}e^{\i p x}e^{\lambda_L(p)|t|}.
\ee
For the specific $\lambda_L(p)$ in (\ref{lambdaL}), this integral can be expressed in terms of the modified Bessel function $-\lambda e^{(2-a)|t|}I_x(a|t|)$. However, we will keep the analysis below somewhat more general, allowing for other functional forms of $\lambda_L(p)$ in \eqref{1d1scramblon}. 

For large $x$ and $t$ (\ref{1d1scramblon}) can be approximated by a saddle point at an imaginary value  $p = \i {\sf P}$:
\be \label{hpt}
\langle \h(x,t)\h(0,0)\rangle \approx -\frac{\lambda}{\sqrt{2\pi \Lambda''({\sf P})t}} e^{-{\sf P} x + \LambdaL({\sf P})|t|}, \hspace{20pt} \frac{x}{t} = \LambdaL'({\sf P}),
\ee
where for convenience we defined
\be
\Lambda(P) \equiv \lambda_L(\i P). \label{lambdadef}
\ee
So we conclude that the scramblon propagator is exponentially decaying in space and exponentially growing in time -- with coefficients that depend on the ratio $x/t$.

\subsection{Loop corrections to the OTOC}
Now, let's consider the OTOC 
\be\label{firsttwoterms}
\langle e^{\Delta_W \h(x,t)} e^{\Delta_V \h(0,0)} \rangle_{\text{exact}} = 1 + \Delta_W\Delta_V \langle \h(x,t)\h(0,0)\rangle + O(\lambda^2).
\ee
where we have introduced the notation $\langle... \rangle_{\text{exact}}$ to refer to expectation values in the full action \eqref{1daction}, and  $\braket{...}$ refers to expectation values in $I_{\text{quad}}$. 
The second term corresponds to single scramblon exchange, which we represent by the following diagram:
\be\label{tree1d}
\begin{tikzpicture}[scale=.7, rotate=0, baseline={([yshift=-0.15cm]current bounding box.center)}]
    \draw[line width=1pt, black,decorate, decoration = {snake, amplitude = 1, segment length = 6pt}] (0,0)  -- (4,4);
    \filldraw[black] (0,0) circle (0.1) node[black, left=4] {$(0,0)$};
    \filldraw[very thick, fill = white] (4,4) circle (0.1) node[black, left=4] {$(x,t)$};
   \draw[very thick, ->] (0,3) -- (0,3.5) node[above] {$t$};
   \draw[very thick, ->] (0,3) -- (.5,3) node[right] {$x$};
\end{tikzpicture}
\ee
The action (\ref{1daction}) has scramblon self-interaction vertices with arbitrary degree, and at order $\lambda^2$, one finds the same set of diagrams that we analyzed in the $(0+1)d$ case in section \ref{sec:loop1}:
\begin{align}\label{oneloopdiags}
\begin{tikzpicture}[scale=.7, rotate=0, baseline={([yshift=-0.15cm]current bounding box.center)}]
    \draw[line width=1pt, black,decorate, decoration = {snake, amplitude = 1, segment length = 6pt}] (0,0) to[out = 30, in = 240] (4,4);
    \draw[line width=1pt, black,decorate, decoration = {snake, amplitude = 1, segment length = 6pt}] (0,0) to[out = 60, in = 210] (4,4);
    \filldraw[black] (0,0) circle (0.1);
    \filldraw[very thick, fill = white] (4,4) circle (0.1);
    \node at (2,-0.5) {(a)};
\end{tikzpicture}\hspace{10pt} + \hspace{10pt} 
\begin{tikzpicture}[scale=.7, rotate=0, baseline={([yshift=-0.15cm]current bounding box.center)}]
    \draw[line width=1pt, black,decorate, decoration = {snake, amplitude = 1, segment length = 5pt}] (0,0) -- (2,2);
    %\draw[very thick, dashed] (2,0) -- (4,4);
    \draw[line width=1pt, black,decorate, decoration = {snake, amplitude = 1, segment length = 5pt}] (2,2) to[out = 20, in = -110] (4,4);
    \draw[line width=1pt, black,decorate, decoration = {snake, amplitude = 1, segment length = 5pt}] (2,2) to[out = 70, in = -160] (4,4);
    \filldraw[black] (0,0) circle (0.1);
    \filldraw[very thick, fill = white] (4,4) circle (0.1);
    \node at (2,-0.5) {(b)};
\end{tikzpicture}\hspace{10pt} + \hspace{10pt} 
\begin{tikzpicture}[scale=.7, rotate=180, baseline={([yshift=-0.15cm]current bounding box.center)}]
   \draw[line width=1pt, black,decorate, decoration = {snake, amplitude = 1, segment length = 5pt}] (0,0) -- (2,2);
    %\draw[very thick, dashed] (2,0) -- (4,4);
    \draw[line width=1pt, black,decorate, decoration = {snake, amplitude = 1, segment length = 5pt}] (2,2) to[out = 20, in = -110] (4,4);
    \draw[line width=1pt, black,decorate, decoration = {snake, amplitude = 1, segment length = 5pt}] (2,2) to[out = 70, in = -160] (4,4);
    \filldraw[very thick, fill = white] (0,0) circle (0.1);
    \filldraw[black] (4,4) circle (0.1);
    \node at (2,4.5) {(c)};
\end{tikzpicture}\hspace{10pt} + \hspace{10pt} 
\begin{tikzpicture}[scale=.7, rotate=0, baseline={([yshift=-0.15cm]current bounding box.center)}]
    \draw[line width=1pt, black,decorate, decoration = {snake, amplitude = 1, segment length = 5pt}] (0,0) -- (1,1);
    \draw[line width=1pt, black,decorate, decoration = {snake, amplitude = 1, segment length = 5pt}] (3,3) -- (4,4);
    %\draw[very thick, dashed] (2,0) -- (4,4);
    \draw[line width=1pt, black,decorate, decoration = {snake, amplitude = 1, segment length = 5pt}] (1,1) to[out = 20, in = -110] (3,3);
    \draw[line width=1pt, black,decorate, decoration = {snake, amplitude = 1, segment length = 5pt}] (1,1) to[out = 70, in = -160] (3,3);
    \filldraw[black] (0,0) circle (0.1);
    \filldraw[very thick, fill = white] (4,4) circle (0.1);
    \node at (2,-0.5) {(d)};
\end{tikzpicture}
\end{align}
In the $(0+1)d$ case, we just evaluated these diagrams explicitly, and found that they became important relative to the tree level answer (\ref{tree1d}) near the scrambling time where the tree level answer itself becomes of order one. In the $(1+1)d$ case, these diagrams are more dangerous, and to gauge their effect we will have to proceed more carefully. 

Let's start by explaining why these diagrams can be larger than in the $(0+1)d$ case. Diagram (a) from (\ref{oneloopdiags}) is proportional to the square of the single-scramblon contribution, so it will be small if the single-scramblon exchange is small, much as in the $(0+1)d$ case. However, diagrams (b), (c), (d) are different. Recall that the scramblon propagator grows exponentially in time and decays exponentially in space. By allocating more of the time interval $[0,t]$ and less of of the space interval $[0,x]$ to the ``loop,'' the overall diagram can become large.

For example, consider the case of diagram (b) from (\ref{oneloopdiags}). Suppose that $x$ and $t$ are both large, but are arranged so that the single-scramblon exchange is small $\langle \h(x,t)\h(0,0)\rangle \ll 1$. If $x$ and $t$ are large enough, it will nevertheless be possible to find intermediate points $(\tilde x,\tilde t)$ such that the propagator connecting this intermediate point to the final point is large $\langle \h(x,t)\h(\tilde x,\tilde t)\rangle \gg 1$. In that situation the following set of diagrams will be out of control, in the sense that adding more loops makes the diagram larger:\footnote{More precisely, for any $x$ and $t$, for sufficiently large $n$ the diagram in \eqref{n_loop} is dominated by an $n$-dependent saddle-point value of $(\tilde x, \tilde t)$ that lies between (0,0) and $(x,t)$. We can check that the saddle-point result for \eqref{n_loop} grows faster in time than $\braket{\h(0;0)\h(t;x)}^n$, and also increases with $n$.}
\begin{align}\label{loopdiag}
\begin{tikzpicture}[scale=.75, rotate=0, baseline={([yshift=-0.15cm]current bounding box.center)}]
    \draw[line width=1pt, black,decorate, decoration = {snake, amplitude = 1, segment length = 6pt}] (0,0)  -- (4,4);
    \filldraw[black] (0,0) circle (0.1) node[black, left=4] {$(0,0)$};
   \filldraw[very thick, fill = white] (4,4) circle (0.1) node[black, left=4] {$(x,t)$};
   \draw[very thick, ->] (0,3) -- (0,3.5) node[above] {$t$};
   \draw[very thick, ->] (0,3) -- (.5,3) node[right] {$x$};
\end{tikzpicture}
\hspace{10pt} + \hspace{10pt}
\begin{tikzpicture}[scale=.75, rotate=0, baseline={([yshift=-0.15cm]current bounding box.center)}]
    \draw[line width=1pt, black,decorate, decoration = {snake, amplitude = 1, segment length = 6pt}] (0,0) -- (3,1);
    %\draw[very thick, dashed] (2,0) -- (4,4);
    \draw[line width=1pt, black,decorate, decoration = {snake, amplitude = 1, segment length = 6pt}] (3,1) to[out = 45, in = -80] (4,4);
    \draw[line width=1pt, black,decorate, decoration = {snake, amplitude = 1, segment length = 6pt}] (3,1) to[out = 100, in = -140] (4,4);
    \filldraw[red]  (3,1) circle (0.1) node[above  = 5, left = 5] {$(\xm,\tm)$};
    \filldraw[black] (0,0) circle (0.1);
    \filldraw[very thick, fill = white] (4,4) circle (0.1);
\end{tikzpicture}
\hspace{10pt} + \hspace{10pt}
\begin{tikzpicture}[scale=.75, rotate=0, baseline={([yshift=-0.15cm]current bounding box.center)}]
    \draw[line width=1pt, black,decorate, decoration = {snake, amplitude = 1, segment length = 6pt}] (0,0) -- (3,1);
    %\draw[very thick, dashed] (2,0) -- (4,4);
    \draw[line width=1pt, black,decorate, decoration = {snake, amplitude = 1, segment length = 6pt}] (3,1) to[out = 45, in = -80] (4,4);
    \draw[line width=1pt, black,decorate, decoration = {snake, amplitude = 1, segment length = 6pt}] (3,1) to[out = 100, in = -140] (4,4);
    \draw[line width=1pt, black,decorate, decoration = {snake, amplitude = 1, segment length = 6pt}] (3,1) to (4,4);
    \filldraw[red]  (3,1) circle (0.1);
    \filldraw[black] (0,0) circle (0.1);
    \filldraw[very thick, fill = white] (4,4) circle (0.1);
\end{tikzpicture} \hspace{10pt} + \hspace{10pt} \cdots
 \end{align}
Here, the diagram means (we will explain the vertex in more detail below)
\be 
\begin{tikzpicture}[scale=.75, rotate=0, baseline={([yshift=-0.15cm]current bounding box.center)}]
    \draw[line width=1pt, black,decorate, decoration = {snake, amplitude = 1, segment length = 6pt}] (0,0) -- (3,1);
    \draw[line width=1pt, black,decorate, decoration = {snake, amplitude = 1, segment length = 6pt}] (3,1) to[out = 45, in = -80] (4,4);
    \draw[line width=1pt, black,decorate, decoration = {snake, amplitude = 1, segment length = 6pt}] (3,1) to[out = 100, in = -140] (4,4);
   % \draw[line width=1pt, black,decorate, decoration = {snake, amplitude = 1, segment length = 6pt}] (3,1) to (4,4);
    \draw[very thick, dotted] (3.9,2.4) -- (3,2.7) node[above=3, right = 4] {$n$};
    \filldraw[red]  (3,1) circle (0.1);
    \filldraw[black] (0,0) circle (0.1);
    \filldraw[very thick, fill = white] (4,4) circle (0.1);
\end{tikzpicture} \hspace{20pt} = \hspace{20pt}  - \frac{2\Delta_V\Delta_W^n}{\lambda n!} \int_0^{t} d \tilde t \int d \tilde x  \braket{ \h(\tilde x,\tilde t)\h(0,0)}  \braket{\h(x,t)\h(\tilde x,\tilde t)}^n  \label{n_loop} 
\ee
Let's emphasize the difference between the $(0+1)d$ and $(1+1)d$ cases. In both cases if $t$ is sufficiently large, the diagrams in (\ref{oneloopdiags}) and (\ref{loopdiag}) will be important. But in the $(0+1)d$ dot, this only happens once the single-scramblon diagram {\it itself} becomes of order one. In the $(1+1)d$ chain, large $x$ can ensure that the single-scramblon diagram is small even though the above set of loop diagrams is large. 

The interpretation is as follows. Loops in the scramblon propagators represent OTOC saturation effects. Even if the overall OTOC is far from saturation, there can be a contribution to the OTOC in which an initial step of scrambling propagates rapidly in space (between the origin and $(\tilde x,\tilde t)$) and then acts as a source for a more relaxed phase of scrambling in which saturation effects are important (between $(\tilde x,\tilde t)$ and $(x,t)$). Diagrams (c) and (d) can be given similar interpretations.

\subsection{Early breakdown of the single-scramblon approximation}\label{BSYKbreak}
After summing all loop diagrams, the OTOC is expected to have the following qualitative behavior. As a function of $t$, it starts out close to one, with a small correction given by the single-scramblon propagator. This correction has a negative sign so it reduces the value of the OTOC. At late times, the OTOC should approach zero. To organize the following discussion, we will define three intermediate times for each $x$:
\begin{align}
t_*(x): &\hspace{15pt} \text{the OTOC $\langle e^{\Delta_W \h(x,t)} e^{\Delta_V \h(0,0)} \rangle_{\text{exact}}$ becomes less than  $\sim\tfrac{1}{2}$}\\
t_1(x) : &\hspace{15pt}\text{the single-scramblon contribution $\langle \h(x,t)\h(0,0)\rangle$ becomes $O(1)$}\\
t_{\text{break}}(x) : &\hspace{15pt} \text{the remainder in (\ref{firsttwoterms}) exceeds the single-scramblon term}
\end{align}
In the $(0+1)d$ dot, all three of these times are approximately equal. In the $(1+1)d$ chain, Xu and Swingle \cite{Xu:2018dfp} show that that $t_{\text{break}} \ll t_1 \ll t_*$, and they conjecture an approximate form for the OTOC by mapping the problem to the noisy FKPP equation \cite{brunet2006phenomenological}. Using the Feynman diagrams, we will not get so far, but we will show that $t_{\text{break}} = t_1 = t_*$ is inconsistent for sufficiently large $x$. The advantage of the Feynman diagram method is that we will also be able to apply it to the non-Brownian low-temperature chain in the next section.

The interaction vertex for the Brownian chain is
\be
I_{\text{int}} = \sum_{i<j}\sigma(i,j)\sum_{\tilde x} \frac{1}{\lambda}\int_0^t \d \tilde t \left[(1-a) e^{\h_{ij}(\tilde x, \tilde t)} + a e^{\frac{\h_{ij}(\tilde x, \tilde t) + \h_{ij}(\tilde x+1, \tilde t)}{2}}\right]_{\text{cubic and higher}}
\ee
Here the subscript ``cubic and higher'' reminds us that the constant, linear, and quadratic terms are not interactions and should be subtracted from $I_{\text{int}}$. To simplify the notation, we will ignore the difference between $\h(\tilde t;\tilde{x})$ and $\h(\tilde t;\tilde{x}+1)$ in the interaction term, and we will replace the sum over $\tilde x$ by an integral.  We can further omit the sum over contours by including a factor of two for each interaction vertex. The reason for this is that because the propagator vanishes in the TOC configuration (\ref{prop_p}), interactions are restricted to pairs of contours $i,j$ form an OTOC configuration with the operators that they connect with in the future and past. There are two such choices of $i<j$ as explained in  \eqref{contourSUMMM}. After these simplifications, the vertex is
\be
I_{\text{int}} = \frac{2}{\lambda} \int_0^t \d \tilde t \int \d \tilde x  \left[e^{\h(\tilde x, \tilde t)} -1-\h(\tilde x,\tilde t) - \frac{\h(\tilde x, \tilde t)^2}{2}\right]. \label{i_int}
\ee

We would now like to sum over all diagrams of the type shown in (\ref{loopdiag}). For this, note that the exact expression for the OTOC can be rewritten as 
\begin{align} \label{326}
&\langle e^{\Delta_W \h(x,t)} e^{\Delta_V \h(0,0)}  \rangle_{\text{exact}} = \braket{e^{-I_{\text{int}}(0, t)}e^{\Delta_W \h(x,t)} e^{\Delta_V \h(0,0)}}  \\
& ~~= \braket{e^{\Delta_W \h(x, t)} e^{\Delta_V \h(0,0)}} - \frac{2}{\lambda} \int_0^t \d \tilde t \int \d \tilde x \Big \langle e^{-I_{\text{int}}(\tilde t, t)}\left[e^{\h(\tilde x, \tilde t)}- 1 - \h( \tilde x, \tilde t) - \frac{\h(\tilde x, \tilde t )^2}{2}\right] e^{\Delta_V \h(0,0)} \Big \rangle   \notag %\nn 
%& ~~= \braket{e^{\Delta_W \h(x, t)} e^{\Delta_V \h(0,0)}} - \sum_{n=0}^{\infty} \frac{\Delta_V^n}{n!}  \frac{2}{\lambda} \int_0^t \d \tilde t \int \d \tilde x \Big \langle e^{-I_{\text{int}}(\tilde t, t)}\left[e^{\h( \tilde x, \tilde t)}- 1 - \h(\tilde x, \tilde t) - \frac{\h(\tilde x, \tilde t)^2}{2}\right] \h(0,0)^n \Big \rangle   
\end{align} 
Here we use the notation $I_{\text{int}}(t_1, t)$ for a more general integral of the form \eqref{i_int} with the lower endpoint replaced with $t_1$. 
In all the loop diagrams of (\ref{loopdiag}), a single scramblon propagates between the origin and the first interaction vertex at $(\tilde x,\tilde t)$.  All such diagrams come from expanding the $e^{\Delta_V \h(0,0)}$ operator in \eqref{326} to linear order and Wick-contracting this factor of $\h(0,0)$ with one factor of $\h(\tilde t, \tilde x)$ from the square brackets. 
The sum of these loop diagrams then gives the following ``error'' term:
\be\label{firstfactorsecondfactora}
\text{error} = -\frac{2\Delta_V}{\lambda}\int \d \tilde x \d \tilde t \ \Big\langle e^{\Delta_W \h(x,t)} \left(e^{\h(\tilde x,\tilde t)}-1-\h(\tilde x,\tilde t) \right) \Big\rangle_{\text{exact}} \Big\langle  \h(\tilde x,\tilde t)\h(0,0)\Big\rangle.
\ee

To estimate the size of (\ref{firstfactorsecondfactora}) we can first fix $\tilde t$ and ask what values of $\tilde x$ give the largest contribution to the integral. This is a tug of war between the first factor, which prefers $\tilde x$ to be larger (closer to $x$) and the second factor, which prefers $\tilde x$ to be smaller (closer to the origin). The first factor involves at least two scramblon propagators, because the term in parentheses starts at quadratic order in $\h$, so it will naively win the tug of war. However, we expect the first factor in (\ref{firstfactorsecondfactora}) to stop increasing rapidly once $\tilde x$ enters ``backward butterfly cone of $(x,t)$,'' defined by
\be
\mathcal{B}^-(x,t) = \{(x',t') \ \text{ s.t. } \ t-t' > t_*(x-x')\}.
\ee
So the first factor in (\ref{firstfactorsecondfactora}) will win the tug of war until $\tilde x$ enters $\mathcal{B}^-(x,t)$, and then it will stop pulling. This means that we can approximate (\ref{firstfactorsecondfactora}) by its contribution near the boundary of $\mathcal{B}^-(x,t)$. Along this locus the integrand reduces (up to $O(1)$ factors represented by ``$\#$'' below) to just the propagator from the origin to $(\tilde x,\tilde t)$
\be\label{inteh}
\text{error} \hspace{10pt} = \hspace{10pt} \frac{\#}{\lambda}\int_{\partial\mathcal{B}^-(x,t)} \hspace{-10pt}\d \tilde t \  \langle  \h(\tilde x,\tilde t)\h(0,0)\rangle\hspace{10pt} = 
\begin{tikzpicture}[scale=.6, rotate=0, baseline={([yshift=-0.15cm]current bounding box.center)}]
 \filldraw[gray!30] (1.2,0) -- (4-.5,2) to[out = 45, in = 140] (4+.5,2) -- (6.8,0) to[white] (1.2,0);
    \draw[line width=1pt, black,decorate, decoration = {snake, amplitude = 1, segment length = 6pt}] (0,0) -- (2.4,1);
    \filldraw[black] (0,0) circle (0.1);% node[black, left=4] {$(0,0)$};
    \filldraw[very thick, fill = white] (4,4) circle (0.1);% node[black, left=4] {$(x,t)$};
   % \node at (4,.75) {$\mathcal{B}^-(x,t)$};
    \filldraw[red] (2.4,1) circle (0.125);% node[left =12, above = 1] {$(\xm(\tm),\tm)$};
\end{tikzpicture}
\ee

The intuition for this formula is that the error is proportional to the probability that an initial fluctuation in the scrambling process can reach far enough in space such that the subsequent phase of scrambling is near saturation.

Let's now estimate (\ref{inteh}) under the assumption that $t_* = t_1$. Then $\partial\mathcal{B}^-(x,t)$ is the locus such that $\langle \h(x,t)\h(\tilde x,\tilde t)\rangle = 1$, which can be determined from the propagator
\begin{align}
\langle \h(x,t)\h(\tilde x,\tilde t)\rangle &\approx -\frac{\lambda}{ \sqrt{2\pi \Lambda''({\sf P})(t-\tilde t)}} \exp\Big[-{\sf P}(x-\tilde x) + \Lambda( {\sf P})(t - \tilde t)\Big]\\
&\approx - \exp\Big[{\sf P}\tilde x - \Lambda( {\sf P})(t_*(x) - t + \tilde t)\Big].\label{usingcond}
\end{align}
Here we take ${\sf P}$ to be the saddle point value for the propagator between the origin and $(x,t)$, which will be a good approximation if $\tilde x \ll x$ and $\tilde t \ll t$. Next, the propagator that appears in (\ref{inteh}) is
\begin{align}
\langle \h(0,0)\h(\tilde x,\tilde t)\rangle &\approx -\lambda\int \frac{\d \tilde P}{2\pi \i}\exp\Big[-\tilde P\tilde x + \Lambda( \tilde P)\tilde t\Big].
\end{align}
This can be restricted to the backwards butterfly cone by setting $\tilde x$ so that (\ref{usingcond}) is equal to one. Substituting this into (\ref{inteh}), we have
\begin{align}\label{inteh2}
\text{error}  &= \# \int\frac{\d \tilde P}{2\pi \i}\int\d \tilde t  \exp\left[-\frac{\tilde P}{{\sf P}}\Lambda({\sf P})(t_*(x)-t+\tilde t) + \Lambda(\tilde P)\tilde t\right].
\end{align}
The saddle point equations are
\be\label{saddle}
\frac{\Lambda(\tilde {\sf P})}{\tilde {\sf P}} = \frac{\Lambda({\sf P})}{{\sf P}}, \hspace{20pt} \left(\frac{{\sf P}}{\Lambda({\sf P})}\Lambda'(\tilde {\sf P}) - 1\right)\tilde t = t_*(x) - t
\ee
and the saddle point approximation for the integral gives
\be\label{errorover1}
\frac{\text{error}}{\text{1 scramblon}} =\#\frac{\exp\Big[-(t_*(x) - t)\left(\Lambda(\tilde {\sf P}) - \Lambda({\sf P})\right)\Big]}{\frac{\Lambda({\sf P})}{{\sf P}} - \Lambda'(\tilde {\sf P})}.
\ee
An obvious try for a solution of the first equation of (\ref{saddle}) would be $\tilde {\sf P} = {\sf P}$. However, when we plug this into the second equation we find that $\tilde t < 0$ --  this solution is not allowed. Instead, the correct saddle is a larger value of $\tilde {\sf P}$ with the same value of $\Lambda(P)/P$:
\be\label{plot}
\includegraphics[width = .38\textwidth]{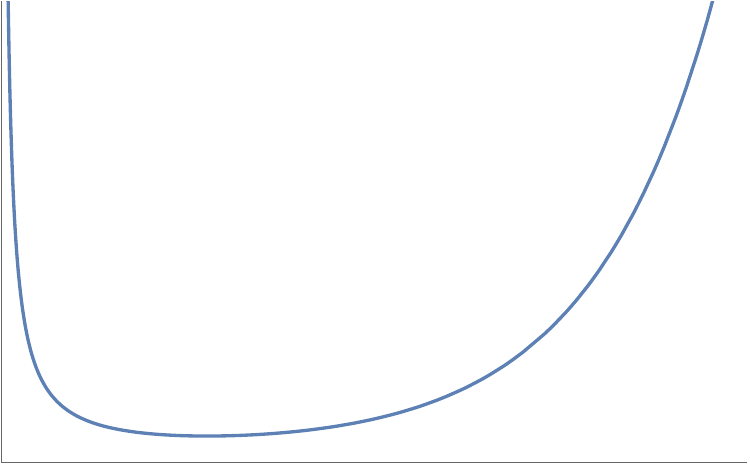} \hspace{40pt}
\def\lkj{6.6}
\def\lkjj{0}
    \begin{tikzpicture}[remember picture, overlay]
         \draw[very thick, gray!40] (-15+\lkj,1.5+\lkjj)--(-8+\lkj,1.5+\lkjj) ;
 \draw[very thick, black,fill=black] (-14.35+\lkj,1.5+\lkjj) circle (0.05) node[left] {${\sf P}$};       
  \draw[very thick, black,fill=black] (-9.4+\lkj,1.5+\lkjj) circle (0.05) node[right] {$\tilde{{\sf P}}$};       
    \draw[very thick, black,fill=black] (-12.8+\lkj,0.25+\lkjj) circle (0.05) node[left] {$P_*$};      
   \node[text width = 60] at (-14.35+\lkj,4+\lkjj)  {$\frac{\Lambda(P)}{P}$};      
      \node[text width = 60] at (-7.4+\lkj,0.2+\lkjj)  {$P$};      
  %          \node[text width = 60] at (1.2,0.2)  {$x/t$};     
  %             \node[text width = 200] at (-2.7,4)  {$\Lambda(\tilde{P}_s) - \Lambda(P_s)$ };    
   \end{tikzpicture}
\ee

We should study this formula in the regime where $t$ is close to $t_*(x)$. To determine ${\sf P}$ and $\tilde {\sf P}$ we can set $t = t_*(x)$. Then 
\be
-{\sf P}x + \Lambda({\sf P}) t_* + \log(\lambda) = 0, \hspace{20pt} \frac{x}{t_*} = \Lambda'({\sf P}),
\ee
where the first equation determines $t_*$ and the second equation then implies
\be
\left({\sf P} - \frac{\Lambda({\sf P})}{\Lambda'({\sf P})}\right) = \frac{\log\frac{1}{\lambda}}{x}.
\ee
For $x \gg \log \frac{1}{\lambda}$, the solution approaches ${\sf P} \to P_*$ where $P_*$ shown in (\ref{plot}) is the value such that $P_*\Lambda'(P_*) = \Lambda(P_*)$. In this limit $\tilde {\sf P}$ will also approach $P_*$, so the coefficient $\Lambda(\tilde {\sf P}) - \Lambda({\sf P})$ in (\ref{errorover1}) will vanish and the denominator will cause a divergence -- the error will exceed the single scramblon contribution well before the scrambling time.\footnote{Note that the approximation $\tilde x \ll x, ~\tilde t \ll t$ used in \eqref{usingcond} is self-consistent as long as 
\be 
\frac{t_{\ast}(x)-t}{\frac{{\sf P}}{\Lambda({\sf P})} \Lambda'(\tilde {\sf P}) -1} \ll t, \quad \quad \frac{\Lambda({\sf P})}{{\sf P}}\left(1+ \frac{1}{\frac{{\sf P}}{\Lambda({\sf P})} \Lambda'(\tilde {\sf P}) -1}\right) (t_{\ast}(x)-t)   \ll x \, . 
\ee
Since we are assuming $x$ and $t$ are both large and considering $t$ close to $t_{\ast}(x)$ in the above discussion, these conditions are satisfied in the regime of interest.}

\section{Ordinary (non-Brownian) large \texorpdfstring{$p$}{p} SYK chain}\label{sec:four}
We will now discuss the case of the ordinary (non-Brownian) SYK chain. Compared to the Brownian case, the scramblon perturbation theory has the following differences:
\begin{enumerate}
\item The interactions involve an integral over two times, not just one.
\item The single-scramblon propagator has a qualitatively different form at low temperatures.
\item The sum over contours for the interaction leads to cancellations at low temperature.
\end{enumerate}
Point 1 does not appear to be qualitatively important, but points 2 and 3 lead to suppression (at low temperature) of the diagrams that led to early breakdown of the single-scramblon approximation in the Brownian case. We are led to speculate that wavefront broadening is much milder in the ordinary chain at low temperature than in the Brownian chain.

\subsection{Scramblon propagator}
The action for the ordinary SYK chain (after setting $\mathcal{J} = 1$) is
\be\label{ordinaryI2}
I = \frac{1}{2\lambda}\sum_x\int_0^\beta \d\tau_1\int_0^\beta \d\tau_2 \left[\frac{1}{4}\partial_1g(\tau_1,\tau_2;x)\partial_2g(\tau_1,\tau_2;x) - (1-a)e^{g(\tau_1,\tau_2;x)} - ae^{\frac{g(\tau_1,\tau_2;x)+g(\tau_1,\tau_2;x+1)}{2}}\right].
\ee
In order to study OTOCs, we will deform the contour between $[0,\beta]$ into a double Keldysh-Schwinger timefold. The saddle point for the dynamical $g$ variable is $g(\tau_1,\tau_2;x) = {\sf g}(\tau_1,\tau_2)$ where
\be\label{saddlenon}
{\sf g}(\tau_1,\tau_2) = 2\log \frac{\cos\frac{\pi v}{2}}{\cos\left[\frac{\pi v}{2}(1 - 2\frac{\tau_{21}}{\beta})\right]}, \hspace{20pt} \frac{\pi v}{\beta} = \cos\frac{\pi v}{2}.
\ee
One can then expand in fluctuations around this saddle
\be
g(\tau_1,\tau_2;x) = {\sf g}(\tau_1,\tau_2) + h(\tau_1,\tau_2;x).
\ee
Propagator for the fluctuation $h$ is not known in closed form except in the limit of large Lorentzian time separation $\text{Im}(\tau_3),\text{Im}(\tau_4) \gg \text{Im}(\tau_1),\text{Im}(\tau_2)$, where it was studied in \cite{gu2019relation}. In momentum space, the resulting propagator is
\be\label{genOTOC}
\lr h(\tau_1,\tau_2;p) h(\tau_3,\tau_4;-p)\rr = -\lambda \frac{\exp\left[\i\lambda_L(p)\frac{\frac{\beta}{2}-\tau_3-\tau_4+\tau_1+\tau_2}{2}\right]}{c(p)} \left[\frac{1}{\cos [\pi v(\frac{1}{2}-\frac{\tau_{21}}{\beta})] \cos[\pi v(\frac{1}{2}-\frac{\tau_{43}}{\beta})]}\right]^{\frac{\beta\lambda_L(p)}{2\pi v}}.
\ee
In this formula, several points require explanation. First, $v$ is a function of temperature defined in (\ref{saddlenon}). Second, $\lambda_L(p)$ is the Lyapunov exponent as a function of momentum and temperature, which can be derived using the method in \cite{gu2019relation}. For large $p$ SYK chain we get (see also Appendix D of \cite{choi2023effective}) :
\be
\lambda_L(p)=\frac{2\pi v}{\beta}\left(\sqrt{\tfrac{9}{4} - a(1-\cos p)} - \frac{1}{2}\right).
\ee
Third, $c(p)$ is a function with the important property that it is proportional to $\cos(\frac{\beta\lambda_L(p)}{4})$, so it has a linear zero at the the location $p = \i {\sf P}_\text{pole}$ defined by the condition $\lambda_L(\i {\sf P}_\text{pole}) = \frac{2\pi}{\beta}$. Finally, the above formula is valid if the arrangement of the points is
\be\label{contourorder}
\text{Re}(\tau_1) < \text{Re}(\tau_3) < \text{Re}(\tau_2) < \text{Re}(\tau_4) < \beta + \text{Re}(\tau_1).
\ee
Other OTOC arrangements can be reduced to this one by relabeling points. 

Let's now discuss the Fourier transform of this propagator back to position space. For example, choosing a simple configuration where the Euclidean parts of the times are equally spaced around the thermal circle, the propagator is
\be
\langle h(0,\tfrac{\beta}{2};0)h(\tfrac{\beta}{4}+\i t,\tfrac{3\beta}{4}+\i t;x)\rangle =
-\lambda\int_{-\pi}^\pi \frac{\d p}{2\pi} \frac{e^{\i p x + \lambda_L(p)t}}{c(p)}.
\ee
Compared to the Brownian chain, a minor difference is that $\lambda_L(p)$ is now a somewhat different function. The major difference is the factor of $1/c(p)$. Recall that in the Brownian case, we analyzed the propagator by deforming the contour for $p$ so that it passed through a saddle point at a positive imaginary value $p = \i {\sf P}$ that depended on $x/t$. In the present case, we can attempt to do the same. However, the zero in $c(p)$ creates a pole in the integrand at a particular imaginary value $p = \i {\sf P}_{\text{pole}}$, and for sufficiently large $x/t$ or sufficiently low temperature ($v$ close to 1), the pole will be closer to the real axis than the saddle point, and the integral will be dominated by the contribution of the pole.

This interplay of the pole and the saddle has been discussed previously in \cite{Shenker:2014cwa,gu2017local,gu2019relation}.

In the regime where the saddle dominates, we believe the model will behave similarly to the Brownian case. So from this point forwards, we will assume that $x/t$ is large enough and/or the temperature is low enough so that the pole dominates, that is, 
\be 
\braket{h(0, \tfrac{\beta}{2};0) h(\tfrac{\beta}{4}+ \i t, \tfrac{3\beta}{4} + \i t; x)} \sim - \lambda e^{- {\sf P_{pole}} \, x + \Lambda({\sf P_{pole}}) \,  t },  
\ee
with $\Lambda$ defined as before in \eqref{lambdadef}. The key difference from the saddle-dominated propagator in $\eqref{hpt}$ is that $\ppole$ does not depend on $x$ and $t$. 
%, and we will use the notation
%\be
%\begin{tikzpicture}[scale=0.5, rotate=0, baseline={([yshift=-0.15cm]current bounding box.center)}]
% \filldraw[black] (-2,0) circle (0.125);
% \draw[line width=1pt, decorate, decoration = {snake, segment length = 7pt, amplitude = 1}] (-2,0) to (2,0);
% \draw[very thick, red, fill = red] (2,0) circle (0.125);
%\end{tikzpicture}\hspace{5 pt}=\langle \h(0,0)\h(x,t)\rangle_{\text{pole}} =  e^{-{\sf P_{pole}}x} \text{Res}\langle h(0,\tfrac{\beta}{2};i{\sf P_{pole}})h(\tfrac{\beta}{4}+\i t,\tfrac{3\beta}{4}+\i t;-i{\sf P_{pole}})\rangle
%\ee

\subsection{Double commutator scramblon propagator}
Before discussing loop effects, it will be useful to consider the double commutator
\begin{align}
&\langle [W(0),V(\i t)][W(\tfrac{\beta}{2}),V(\tfrac{\beta}{2}+\i t)]\rangle\\ \notag
&\hspace{20pt}=\hspace{20pt}\begin{tikzpicture}[scale=0.7, rotate=0, baseline={([yshift=-0.15cm]current bounding box.center)}]
  \def\xc{3}
  \def\yy{.3}
  \def\yyy{.75}
  \draw[very thick, gray!50, rounded corners=.5mm] (\xc,0) -- (0,0) -- (0,-\yy) -- (\xc,-\yy) -- (\xc,-\yyy - \yy) -- (0,-\yyy - \yy) -- (0,-\yyy - 2*\yy) -- (\xc,-\yyy - 2*\yy) --  (\xc,-2*\yy - 2*\yyy);
  \filldraw[black] (0,-.5*\yy) circle (0.125);
  \filldraw[black] (0,-1*\yyy-1.5*\yy) circle (0.125);
  \draw[very thick, black, fill = white] (\xc,-1*\yy) circle (0.125);
  \draw[very thick, black, fill = white] (\xc,-1*\yyy-2*\yy) circle (0.125); 
\end{tikzpicture} \hspace{20pt} - \hspace{20pt} \begin{tikzpicture}[scale=0.7, rotate=0, baseline={([yshift=-0.15cm]current bounding box.center)}]\label{sumcontour}
  \def\xc{3}
  \def\yy{.3}
  \def\yyy{.75}
  \draw[very thick, gray!50, rounded corners=.5mm] (\xc,0) -- (0,0) -- (0,-\yy) -- (\xc,-\yy) -- (\xc,-\yyy - \yy) -- (0,-\yyy - \yy) -- (0,-\yyy - 2*\yy) -- (\xc,-\yyy - 2*\yy) --  (\xc,-2*\yy - 2*\yyy);
  \filldraw[black] (0,-.5*\yy) circle (0.125);
  \filldraw[black] (0,-1*\yyy-1.5*\yy) circle (0.125);
  \draw[very thick, black, fill = white] (\xc,0*\yy) circle (0.125);
  \draw[very thick, black, fill = white] (\xc,-1*\yyy-2*\yy) circle (0.125); 
\end{tikzpicture} \hspace{20pt} + \hspace{20pt}
\begin{tikzpicture}[scale=0.7, rotate=0, baseline={([yshift=-0.15cm]current bounding box.center)}]
  \def\xc{3}
  \def\yy{.3}
  \def\yyy{.75}
  \draw[very thick, gray!50, rounded corners=.5mm] (\xc,0) -- (0,0) -- (0,-\yy) -- (\xc,-\yy) -- (\xc,-\yyy - \yy) -- (0,-\yyy - \yy) -- (0,-\yyy - 2*\yy) -- (\xc,-\yyy - 2*\yy) --  (\xc,-2*\yy - 2*\yyy);
  \filldraw[black] (0,-.5*\yy) circle (0.125);
  \filldraw[black] (0,-1*\yyy-1.5*\yy) circle (0.125);
  \draw[very thick, black, fill = white] (\xc,0*\yy) circle (0.125);
  \draw[very thick, black, fill = white] (\xc,-1*\yyy-1*\yy) circle (0.125); 
\end{tikzpicture} \hspace{20pt} - \hspace{20pt}
\begin{tikzpicture}[scale=0.7, rotate=0, baseline={([yshift=-0.15cm]current bounding box.center)}]
  \def\xc{3}
  \def\yy{.3}
  \def\yyy{.75}
  \draw[very thick, gray!50, rounded corners=.5mm] (\xc,0) -- (0,0) -- (0,-\yy) -- (\xc,-\yy) -- (\xc,-\yyy - \yy) -- (0,-\yyy - \yy) -- (0,-\yyy - 2*\yy) -- (\xc,-\yyy - 2*\yy) --  (\xc,-2*\yy - 2*\yyy);
  \filldraw[black] (0,-.5*\yy) circle (0.125);
  \filldraw[black] (0,-1*\yyy-1.5*\yy) circle (0.125);
  \draw[very thick, black, fill = white] (\xc,-1*\yy) circle (0.125);
  \draw[very thick, black, fill = white] (\xc,-1*\yyy-1*\yy) circle (0.125); 
\end{tikzpicture}
\end{align}
The exponentially growing pieces are contained in the two OTOC cases (the first and third diagrams). The contribution of the scramblon propagator to these pieces is proportional to 
\begin{align}
&\langle h(0,\tfrac{\beta}{2};0)h(\epsilon+\i t,\tfrac{\beta}{2}+\epsilon+\i t;x)\rangle + \langle h(0,\tfrac{\beta}{2};0)h(\tfrac{\beta}{2}-\epsilon+\i t,\beta-\epsilon+\i t;x)\rangle\\
&\hspace{40pt}=
-\lambda\int_{-\pi}^\pi \frac{\d p}{2\pi} \frac{e^{\i p x + \lambda_L(p)t}}{c(p)}\left[e^{\i\lambda_L(p)\frac{\beta}{4}} + e^{-\i\lambda_L(p)\frac{\beta}{4}}\right].
\end{align}
The term in brackets has a linear zero that cancels the zero in $c(p)$, meaning that the integrand does not have a pole at $p = \i P_\text{pole}$.

The interpretation of this is that in the pole-dominated regime, the leading contribution to the OTOC cancels out when we compute the commutator OTOC. This doesn't mean that the commutator OTOC vanishes, but it means that it is always dominated by the saddle point, even in the regime where the pole contribution to the scramblon propagator dominates over the saddle point \cite{Gu:2021xaj}.

\subsection{Loop corrections to the OTOC}
We would like to check if the dangerous loop corrections analogous to (\ref{loopdiag}) still cause an early breakdown of the single scramblon approximation. We will consider sufficiently large $x$ and/or low enough temperature so that the single scramblon contribution is dominated by the pole. As we mentioned in point 1 at the beginning of the section, in principle the interactions involve an integral over two times. However the interactions will be almost local in time, partly because \ref{genOTOC} decays rapidly when $\text{Im}(\tau_4-\tau_3)$ is large. So in the rough argument below, we will approximate the interactions as happening at a single time.

Let's start with the correction
\be\label{Startwith}
\begin{tikzpicture}[scale=.7, rotate=0, baseline={([yshift=-0.15cm]current bounding box.center)}]
    \draw[line width=1pt, black,decorate, decoration = {snake, amplitude = 1, segment length = 5pt}] (0,0) -- (2,2);
    %\draw[very thick, dashed] (2,0) -- (4,4);
    \draw[line width=1pt, black,decorate, decoration = {snake, amplitude = 1, segment length = 5pt}] (2,2) to[out = 20, in = -110] (4,4);
    \draw[line width=1pt, black,decorate, decoration = {snake, amplitude = 1, segment length = 5pt}] (2,2) to[out = 70, in = -160] (4,4);
    \filldraw[black] (0,0) circle (0.125);
     \filldraw[red] (2,2) circle (0.125)  node[above  = 5, left = 5] {$(\xm,\tm)$};
    \filldraw[very thick, fill = white] (4,4) circle (0.125);
\end{tikzpicture}
\ee
The sum over contours (\ref{contourSUMMM}) reduces to two terms
\begin{align}
& \begin{tikzpicture}[scale=0.7, rotate=0, baseline={([yshift=-0.15cm]current bounding box.center)}]\label{sumcontour}
  \def\xc{5}
  \def\yy{.2}
  \def\yyy{1}
  \draw[very thick, gray!50, rounded corners=.5mm] (\xc,0) -- (0,0) -- (0,-\yy) -- (\xc,-\yy) -- (\xc,-\yyy - \yy) -- (0,-\yyy - \yy) -- (0,-\yyy - 2*\yy) -- (\xc,-\yyy - 2*\yy) --  (\xc,-2*\yy - 2*\yyy);
  \filldraw[black] (0,-.5*\yy) circle (0.125);
  \filldraw[black] (0,-1*\yyy-1.5*\yy) circle (0.125);
  \filldraw[red] (\xc/2,0) circle (0.125); %node[above=2] {$-\epsilon + \i t'$};
  \filldraw[red] (\xc/2,-1*\yyy-1*\yy) circle (0.125);
  \draw[very thick, black, fill = white] (\xc,-.5*\yyy-\yy) circle (0.125); %node[anchor=west] {$\theta_5$};
  \draw[very thick, black, fill = white] (\xc,-1.5*\yyy-2*\yy) circle (0.125); %node[anchor=west] {$\theta_6$};
\end{tikzpicture} \hspace{20pt} + \hspace{20pt}
\begin{tikzpicture}[scale=0.7, rotate=0, baseline={([yshift=-0.15cm]current bounding box.center)}]
  \def\xc{5}
  \def\yy{.2}
  \def\yyy{1}
  \draw[very thick, gray!50, rounded corners=.5mm] (\xc,0) -- (0,0) -- (0,-\yy) -- (\xc,-\yy) -- (\xc,-\yyy - \yy) -- (0,-\yyy - \yy) -- (0,-\yyy - 2*\yy) -- (\xc,-\yyy - 2*\yy) --  (\xc,-2*\yy - 2*\yyy);
  \filldraw[black] (0,-.5*\yy) circle (0.125);
  \filldraw[black] (0,-1*\yyy-1.5*\yy) circle (0.125);
  \filldraw[red] (\xc/2,-1*\yy) circle (0.125); %node[above=4] {$\epsilon + \i t'$};
  \filldraw[red] (\xc/2,-1*\yyy-2*\yy) circle (0.125);
  \draw[very thick, black, fill = white] (\xc,-.5*\yyy-\yy) circle (0.125); %node[anchor=west] {$\theta_5$};
  \draw[very thick, black, fill = white] (\xc,-1.5*\yyy-2*\yy) circle (0.125); %node[anchor=west] {$\theta_6$};
\end{tikzpicture}
\end{align}
In the Brownian models, these two contour configurations combine to give a factor of two. However, more generally what happens is the sum converts the propagator between the origin and the interaction point into a double-commutator scramblon propagator. This means that the leading (pole dominated) contribution cancels between the two diagrams.

We can try to interpret this cancellation using the language of coherent and incoherent scrambling, as discussed by Gu, Kitaev, and Zhang in section 4.2 of \cite{Gu:2021xaj}. The double commutator corresponds to the {\it probablity} of a type of scrambling taking place, and it receives contributions only from the ``incoherent'' part of the scramblon propagator. On the other hand, the ordinary OTOC also receives contributions from an interference effect that corresponds to the {\it amplitude} for a type of scrambling to take place. The pole part of the scramblon propagator corresponds to purely coherent scrambling -- to an amplitude rather than a probability. Now, let's consider the diagrams (\ref{Startwith}). In the Brownian (or high temperature) model we interpret these as representing an essentially classical stochastic effect where the initial propagator to $(\tilde x, \tilde t)$ provides a  source for later scrambling based from that point. Such an effect would be included in the noisy FKPP equation. It seems reasonable that the part of the initial propagator that represents an amplitude rather than a probability does not contribute to such an effect.

Of course, the cancellation doesn't mean that (\ref{Startwith}) vanishes exactly, it just means that we need to replace the initial propagator with the incoherent double-commutator propagator. More generally, the diagrams (\ref{loopdiag}) become 
\begin{align}\label{loopdiag2}
\begin{tikzpicture}[scale=.75, rotate=0, baseline={([yshift=-0.15cm]current bounding box.center)}]
    \draw[line width=1pt, black,decorate, decoration = {snake, amplitude = 1, segment length = 6pt}] (0,0)  -- (4,4);
    \filldraw[black] (0,0) circle (0.1) node[black, left=4] {$(0,0)$};
   \filldraw[very thick, fill = white] (4,4) circle (0.1) node[black, left=4] {$(x,t)$};
   \draw[very thick, ->] (0,3) -- (0,3.5) node[above] {$t$};
   \draw[very thick, ->] (0,3) -- (.5,3) node[right] {$x$};
\end{tikzpicture}
\hspace{10pt} + \hspace{10pt}
\begin{tikzpicture}[scale=.7, rotate=0, baseline={([yshift=-0.15cm]current bounding box.center)}]
    \draw[line width=1pt, black,decorate, decoration = {snake, amplitude = 1, segment length = 5pt}] (0,0) -- (2,2);
    %\draw[very thick, dashed] (2,0) -- (4,4);
    \draw[line width=1pt, black,decorate, decoration = {snake, amplitude = 1, segment length = 5pt}] (2,2) to[out = 20, in = -110] (4,4);
    \draw[line width=1pt, black,decorate, decoration = {snake, amplitude = 1, segment length = 5pt}] (2,2) to[out = 70, in = -160] (4,4);
    \filldraw[black] (0,0) circle (0.125);
     \filldraw[red] (2,2) circle (0.125)  node[above  = 5, left = 5] {$(\xm,\tm)$};
    \filldraw[very thick, fill = white] (4,4) circle (0.125);
    \DrawBracket{1}{1}{45};
\end{tikzpicture}
\hspace{10pt} + \hspace{10pt}
\begin{tikzpicture}[scale=.7, rotate=0, baseline={([yshift=-0.15cm]current bounding box.center)}]
    \draw[line width=1pt, black,decorate, decoration = {snake, amplitude = 1, segment length = 5pt}] (0,0) -- (2,2);
    %\draw[very thick, dashed] (2,0) -- (4,4);
    \draw[line width=1pt, black,decorate, decoration = {snake, amplitude = 1, segment length = 5pt}] (2,2) to[out = 20, in = -110] (4,4);
    \draw[line width=1pt, black,decorate, decoration = {snake, amplitude = 1, segment length = 5pt}] (2,2) to[out = 70, in = -160] (4,4);
    \draw[line width=1pt, black,decorate, decoration = {snake, amplitude = 1, segment length = 5pt}] (2,2) to (4,4);
    \filldraw[black] (0,0) circle (0.125);
    \filldraw[red] (2,2) circle (0.125);
    \filldraw[very thick, fill = white] (4,4) circle (0.125);
    \DrawBracket{1}{1}{45};
\end{tikzpicture} \hspace{10pt} + \hspace{10pt} \cdots
 \end{align}
where we are using the notation of a propagator with a bracket $\begin{tikzpicture}[scale=.75,rotate=0, baseline={([yshift=-0.15cm]current bounding box.center)}]\DrawBracket{0}{0}{0};\end{tikzpicture}$ to indicate the double-commutator propagator, dominated by the saddle point. Following the same logic as in the Brownian case, we expect that the error in the single scramblon approximation from these diagrams is
\be\label{ratioerrorreg}
\text{error}\hspace{10pt}=\hspace{10pt}\frac{\#}{\lambda}\int_{\partial\mathcal{B}^-}\langle h(0,\tfrac{\beta}{2};0)h(\i \tilde t, \tfrac{\beta}{2}+ \i \tilde t; x)\rangle_{\text{saddle}}\hspace{10pt}= \hspace{10pt} \begin{tikzpicture}[scale=.6, rotate=0, baseline={([yshift=-0.15cm]current bounding box.center)}]
 \filldraw[gray!30] (1.2,0) -- (4-.5,2) to[out = 45, in = 140] (4+.5,2) -- (6.8,0) to[white] (1.2,0);
    \draw[line width=1pt, black,decorate, decoration = {snake, amplitude = 1, segment length = 6pt}] (0,0) -- (2.4,1);
    \filldraw[black] (0,0) circle (0.1);% node[black, left=4] {$(0,0)$};
    \filldraw[very thick, fill = white] (4,4) circle (0.1);% node[black, left=4] {$(x,t)$};
   % \node at (4,.75) {$\mathcal{B}^-(x,t)$};
    \filldraw[red] (2.4,1) circle (0.125);% node[left =12, above = 1] {$(\xm(\tm),\tm)$};
    \DrawBracket{1.25}{.5}{22};
\end{tikzpicture}
\ee
The key difference of this calculation compared to the one in Sec.~\ref{BSYK1d} is that while the propagator from $(0,0)$ to $(\tilde{x},\tilde{t})$ is given by the saddle point, the past butterfly cone $\partial \mathcal{B}_-$  is now defined by the pole contribution. This pole contribution propagates faster in space, so the saddle point propagator has trouble catching up, unless the origin is very close to the butterfly cone. More precisely, following the same logic as in the previous section, one now finds the estimate 
\be\label{}
\frac{\text{error}}{\text{1 scramblon}}=\# \exp\left[-(t_*(x)-t)\left(\LambdaL({\sf \tilde{P}})-\LambdaL({\sf P_{pole}})\right) \right]
\ee
where ${\sf \tilde{P}}$ is the solution with ${\sf \tilde{P}} > {\sf P_\text{pole}}$ to the equation
\be\label{}
\frac{\LambdaL({\sf \tilde{P}})}{{\sf \tilde{P}}} = \frac{\LambdaL({\sf P_{pole}})}{{\sf P_{pole}}}.
\ee
Since ${\sf \tilde{P}}$ and  ${\sf P_\text{pole}}$ are some fixed values (independent of $x$) with $\LambdaL({\sf \tilde{P}}) > \LambdaL({\sf {P}_{pole}})$, we find that the error is suppressed for $t < t_*(x)$ even if $x$ is very large. So in the low temperature regime of the ordinary SYK chain, we do not find evidence for the dramatic breakdown of the single scramblon approximation observed in the Brownian chain with $x \gg \log (1/\lambda)$.

However, we do expect a less-dramatic breakdown of the single-scramblon approximation when $x$ is much larger, $x\sim 1/\lambda$. This is because there are other loop diagrams for which the pole contribution does not cancel out. For example, consider
\be\label{xSYKgroup}
\begin{tikzpicture}[scale=.75, rotate=0, baseline={([yshift=-0.15cm]current bounding box.center)}]
    \draw[line width=1pt, black,decorate, decoration = {snake, amplitude = 1, segment length = 6pt}] (0,0) to[out = 10, in = 260] (2,2);
    \draw[line width=1pt, black,decorate, decoration = {snake, amplitude = 1, segment length = 6pt}] (0,0) to[out = 80, in = 190] (2,2);
     \draw[line width=1pt, black,decorate, decoration = {snake, amplitude = 1, segment length = 6pt}] (2,2) to[out = 10, in = 260] (4,4);
    \draw[line width=1pt, black,decorate, decoration = {snake, amplitude = 1, segment length = 6pt}] (2,2) to[out = 80, in = 190] (4,4);
    \filldraw[black] (0,0) circle (0.1);
    \filldraw[very thick, fill = white] (4,4) circle (0.1);
    %  \filldraw[red] (2,2) circle (0.1);% node[anchor=north west] {$(\tilde{x},\tilde{t})$};
\end{tikzpicture} %\hspace{20pt} + \hspace{20pt} \begin{tikzpicture}[scale=.75, rotate=0, baseline={([yshift=-0.15cm]current bounding box.center)}]
   % \draw[line width=1pt, black,decorate, decoration = {snake, amplitude = 1, segment length = 6pt}] (0,0) to[out = 20, in = 250] (4,4);
   % \draw[line width=1pt, black,decorate, decoration = {snake, amplitude = 1, segment length = 6pt}] (0,0) to[out = 70, in = 200] (4,4);
   % \filldraw[black] (0,0) circle (0.1);
   % \filldraw[very thick, fill = white] (4,4) circle (0.1);
   % \draw[very thick, black] (2-.52,2+.52) to (2+.52,2-.52);
     %  \filldraw[red] (2,2) circle (0.1);% node[anchor=north west] {$(\tilde{x},\tilde{t})$};
%\end{tikzpicture}
\ee
In this case one can use the pole contribution for all of the propagators. The location of the interaction vertex becomes a zero mode, and the diagram is of order
\be
 \lambda \cdot (\text{vol of zero mode})\cdot \left(e^{-\frac{2\pi}{\beta}(t_*(x) - t)}\right)^2.
\ee
The last factor comes from the two scramblon propagators connecting the origin to $(x,t)$, and we have used that $\Lambda({\sf P}_\text{pole}) = \frac{2\pi}{\beta}$. Naively, the volume of the zero mode would be $xt$, but because of OTOC saturation effects, we expect it will be limited to the strip region between the future butterfly cone of $(0,0)$ and the past butterfly cone of $(x,t)$:
\be
\begin{tikzpicture}[scale=.75, rotate=0, baseline={([yshift=-0.15cm]current bounding box.center)}]
      \filldraw[gray!30] (-3,4) -- (0-.5,1+.75) to[out = -50, in = -135] (0+.5,1+.75) -- (3,4) to[white] (-2,4);
    \filldraw[gray!30] (1,0) -- (4-.5,3-.75) to[out = 50, in = 135] (4+.5,3-.75) -- (7,0) to[white] (2,0);
    \filldraw[black] (0,0) circle (0.1) node[black, left=4] {$(0,0)$};
    \filldraw[black] (4,4) circle (0.1) node[black, right=4] {$(x,t)$};
    %\node[text width = 60] at (5,1) {$\mathcal{B}^-$};
    %    \node[text width = 60] at (1,3) {$\mathcal{B}^+$};
\end{tikzpicture}
\ee
The long direction of this strip is proportional to $t$, so very schematically we expect a contribution of order 
\be\label{analogous}
\lambda \cdot t\cdot \left(e^{-\frac{2\pi}{\beta}(t_*(x) - t)}\right)^2.
\ee
This is suppressed by $\lambda t$ relative to the two-scramblon contribution, as in the correction (\ref{0+1dcross}) studied in the $(0+1)d$ model. An important difference in the present case is that $\lambda t$ can be large near the scrambling time $t_*(x)$ if $x$ is sufficiently large, $x\sim 1/\lambda$. It would be interesting to study these diagrams in more detail.\footnote{One also expects $\lambda t$ corrections for a single scramblon propagator and important contributions from coupling between the scramblon and energy density fluctuations.}

\section{Discussion}

Our main motivation was to understand the origin of front-broadening from the perspective of interacting scramblons. We explored a precursor to this effect, which is the breakdown of the single-scramblon approximation well before the scrambling time. We established this breakdown for the Brownian or high temperature SYK chains at large $x$. In the low-temperature case we found a cancellation of the leading diagrams that cause this breakdown, so we expect that front-broadening is much milder at low temperature than it is at high temperatures or in the Brownian model.

In the Brownian (or high temperature) chain, scrambling can be understood in terms of a classical stochastic model \cite{Xu:2018dfp}. For example, one can consider an epidemic in which infected agents move by a random walk process and infect nearby agents until saturation in the local population. This type of process is described at long distances by a noisy FKPP equation \cite{brunet2006phenomenological}. The breakdown in the single scramblon approximation  is due to the possibility of a fluctuation in which a single scramblon (chain of infection) initially propagates rapidly in space, and then acts as a source for a nonlinear period of scrambling in which saturation effects are important. 

At low temperature, the scrambling is coherent in the sense that the scramblon propagator corresponds to an amplitude rather than a probability, and we found a cancellation in the diagrams that led to the breakdown of the single-scramblon approximation in the Brownian case. In this regime we do not expect the classical stochastic model to apply, and it would be nice to find a simple picture that replaces it.

Although we have not considered holographic theories in this paper, it is interesting to try to extrapolate to that case. In holographic theories, the scramblon corresponds to the Pomeron operator that is exchanged in high energy scattering near the black hole horizon. The coherent regime corresponds to dominance by the graviton pole, and the resummation of multi-scramblon exchange diagrams corresponds \cite{Shenker:2014cwa} to the eikonal resummation for gravity \cite{tHooft:1987vrq,Amati:1987uf,Verlinde:1991iu,Kabat:1992tb}. Corrections to the leading eikonal approximation have been studied for flat-space scattering of gravitons in \cite{Amati:1990xe}, and multiplicative corrections of order $G_N \log(s)$ exist (see (5.26) of \cite{Amati:1990xe}), which translates to $\lambda t$ in our case. This could be analogous to the correction in (\ref{analogous}). In the incoherent stringy regime, one might expect a larger breakdown of the eikonal resummation and it seems important to study this further.

\section*{Acknowledgements} 

We thank Zhenbin Yang and Pengfei Zhang for useful discussions at an early stage of this project. This work was supported in part by DOE grant DE-SC0021085, by the Sloan Foundation, and by a grant from the Simons foundation (926198, DS).  SV is supported by Google.

\appendix

\section{Chord methods for large $p$ Brownian SYK}
For a review of the chord diagram method for computing correlation functions in ordinary large $p$ SYK, see \cite{berkooz2019towards,lin2022bulk,Lin:2023trc}. The difference in the Brownian case is that the Hamiltonian chords can only connect points at the same time -- although these points could be on different contours. 

For example, let's first consider the case with no matter chords, only Hamiltonian chords. We will study the computation of $\text{tr}(U(t) U^\dagger(t))$, which involves a Schwinger-Keldysh contour with one time fold:
\be
\begin{tikzpicture}[baseline={(current bounding box.center)}]
    \draw[thick](0,0) -- (3,0) arc (90:-90:.15) -- (0,-.3) arc (270:90:.15);
    \draw (0,0) node[above] {$0$};
    \draw (3,0) node[above] {$t$};
\end{tikzpicture}
\ee
Hamiltonian chords can either connect a contour to itself or to the other contour:
\be
\begin{tikzpicture}[baseline={(current bounding box.center)}]
    \draw[thick](0,0) -- (3,0) arc (90:-90:.15) -- (0,-.3) arc (270:90:.15);
    \draw (0,0) node[above] {$0$};
    \draw (3,0) node[above] {$t$};
    \draw[very thick] (.5,0) -- (.5,-.3);
    \draw[very thick] (.7,0) -- (.7,-.3);
    \draw[very thick] (1.7,0) -- (1.7,-.3);
    \draw[very thick] (2.3,0) -- (2.3,-.3);
    \draw[very thick] (1.3,0) arc (180:360:.1);
    \draw[very thick] (2.4,0) arc (180:360:.1);
    \draw[very thick] (1,-.3) arc (180:0:.1);
\end{tikzpicture}
\ee
Here the chords can only connect at equal times. This was drawn faithfully for the chords that link one side of the contour to the other, but it was not quite drawn accurately for the chords that link a given side to itself. The chord rules for this case are
\begin{align}\label{chordrule2pt}
\text{chord linking same side: } -\frac{1}{2\lambda}\\
\text{chord linking opposite sides: } \frac{1}{\lambda}.
\end{align}
The sum over all such chords gives a time-independent answer, as required by unitarity:
\be
\langle 1\rangle = \left(\sum_{m = 0}^\infty  \frac{t^m}{m!}\frac{1}{\lambda^m}\right) \times \left(\sum_{m = 0}^\infty  \frac{t^m}{m!}\left(\frac{-1}{2\lambda}\right)^m\right)^2 = 1.
\ee

Now, we consider the case of a two point function, where the contour looks like
\be\label{2ptchord}
\begin{tikzpicture}[baseline={(current bounding box.center)}]
    \draw[thick](0,0) -- (3,0) arc (90:-90:.15) -- (0,-.3) arc (270:90:.15);
    \draw[very thick, blue] (0,0) -- (3,0) ;
    \filldraw[blue] (0,0) circle (.075) node[black, above] {$0$};
    \filldraw[blue] (3,0) circle (.075) node[black, above] {$t$};
\end{tikzpicture}
\ee
Now when we sum over chords, we get a factor of $q^\Delta$ for each chord that links the opposite sides, because these chords link with the blue ``matter'' chord. Here we are using the notation
\be
q = e^{-\lambda}.
\ee
The result is 
\be
\langle e^{\Delta g(t,0)}\rangle = \left(\sum_{m = 0}^\infty  \frac{t^m q^{\Delta m}}{m!}\frac{1}{\lambda^m}\right) \times \left(\sum_{m = 0}^\infty  \frac{t^m}{m!}\left(\frac{-1}{2\lambda}\right)^m\right)^2 =  e^{ -\frac{1-q^{\Delta}}{\lambda}t}.
\ee

\subsection{Four point function}\label{usemethod}
Now we discuss the four point function $\langle e^{\Delta_1 g(0,0)}e^{\Delta_2 g(t,t)}\rangle_{\text{OTOC}}$. For this we need an OTOC contour
\be
\begin{tikzpicture}[baseline={(current bounding box.center)}]
    \draw[thick](0,0) -- (3,0) arc (90:-90:.15) -- (0,-.3) arc (90:270:.15) -- (3,-.6) arc (90:-90:.15) -- (0,-.9) arc (270:90:.45);
\end{tikzpicture}
\ee
Let's start by considering the case without matter insertions. We sum over chords with the following rules. Chords connecting contours going the ``same direction'' come with minus signs
\begin{align}
\begin{tikzpicture}[baseline={(current bounding box.center)}]
    \draw[thick](0,0) -- (1,0);
    \draw[thick](0,-.3) -- (1,-.3);
    \draw[thick](0,-.6) -- (1,-.6);
    \draw[thick](0,-.9) -- (1,-.9);
    \draw[very thick] (.5,0) arc (180:360:.1);
\end{tikzpicture}
\hspace{15pt}
\begin{tikzpicture}[baseline={(current bounding box.center)}]
    \draw[thick](0,0) -- (1,0);
    \draw[thick](0,-.3) -- (1,-.3);
    \draw[thick](0,-.6) -- (1,-.6);
    \draw[thick](0,-.9) -- (1,-.9);
    \draw[very thick] (.5,-.3) arc (180:360:.1);
\end{tikzpicture}
\hspace{15pt}
\begin{tikzpicture}[baseline={(current bounding box.center)}]
    \draw[thick](0,0) -- (1,0);
    \draw[thick](0,-.3) -- (1,-.3);
    \draw[thick](0,-.6) -- (1,-.6);
    \draw[thick](0,-.9) -- (1,-.9);
    \draw[very thick] (.5,-.6) arc (180:360:.1);
\end{tikzpicture}
\hspace{15pt}
\begin{tikzpicture}[baseline={(current bounding box.center)}]
    \draw[thick](0,0) -- (1,0);
    \draw[thick](0,-.3) -- (1,-.3);
    \draw[thick](0,-.6) -- (1,-.6);
    \draw[thick](0,-.9) -- (1,-.9);
    \draw[very thick] (.5,-.9) arc (180:360:.1);
\end{tikzpicture}
\hspace{15pt} &= \hspace{15pt} -\frac{1}{2\lambda}
\label{nolinking}
\end{align}
\be
\begin{tikzpicture}[baseline={(current bounding box.center)}]
    \draw[thick](0,0) -- (1,0);
    \draw[thick](0,-.3) -- (1,-.3);
    \draw[thick](0,-.6) -- (1,-.6);
    \draw[thick](0,-.9) -- (1,-.9);
    \draw[very thick] (.5,0) -- (.5,-.6);
\end{tikzpicture}
\hspace{15pt}
\begin{tikzpicture}[baseline={(current bounding box.center)}]
    \draw[thick](0,0) -- (1,0);
    \draw[thick](0,-.3) -- (1,-.3);
    \draw[thick](0,-.6) -- (1,-.6);
    \draw[thick](0,-.9) -- (1,-.9);
    \draw[very thick] (.5,-.3) -- (.5,-.9);
\end{tikzpicture} \hspace{15pt} = \hspace{15pt} -\frac{1}{\lambda}
\ee
Chords linking contours going in opposite directions come with a positive sign:
\be
\begin{tikzpicture}[baseline={(current bounding box.center)}]
    \draw[thick](0,0) -- (1,0);
    \draw[thick](0,-.3) -- (1,-.3);
    \draw[thick](0,-.6) -- (1,-.6);
    \draw[thick](0,-.9) -- (1,-.9);
    \draw[very thick] (.5,0) -- (.5,-.3);
\end{tikzpicture}
\hspace{15pt}
\begin{tikzpicture}[baseline={(current bounding box.center)}]
    \draw[thick](0,0) -- (1,0);
    \draw[thick](0,-.3) -- (1,-.3);
    \draw[thick](0,-.6) -- (1,-.6);
    \draw[thick](0,-.9) -- (1,-.9);
    \draw[very thick] (.5,-.6) -- (.5,-.9);
\end{tikzpicture}
\hspace{15pt}
\begin{tikzpicture}[baseline={(current bounding box.center)}]
    \draw[thick](0,0) -- (1,0);
    \draw[thick](0,-.3) -- (1,-.3);
    \draw[thick](0,-.6) -- (1,-.6);
    \draw[thick](0,-.9) -- (1,-.9);
    \draw[very thick] (.5,-.3) -- (.5,-.6);
\end{tikzpicture}
\hspace{15pt}
\begin{tikzpicture}[baseline={(current bounding box.center)}]
    \draw[thick](0,0) -- (1,0);
    \draw[thick](0,-.3) -- (1,-.3);
    \draw[thick](0,-.6) -- (1,-.6);
    \draw[thick](0,-.9) -- (1,-.9);
    \draw[very thick] (.5,0) -- (.5,-.9);
\end{tikzpicture}\hspace{15pt} = \hspace{15pt} \frac{1}{\lambda}
\ee

In addition to these signs, we have to be careful with the linking numbers of various chords. The chords in (\ref{nolinking}) do not link with any other chords, so their contribution will be simple. However, the other types of chords can link with each other. Let's give a couple of examples. First, consider
\begin{align}
\begin{tikzpicture}[baseline={([yshift=-0.1cm]current bounding box.center)}]
    \draw[thick](0,0) -- (3,0) arc (90:-90:.15) -- (0,-.3) arc (90:270:.15) -- (3,-.6) arc (90:-90:.15) -- (0,-.9) arc (270:90:.45);
    \draw[very thick, blue] (1,-.3) -- (1,-.6);
    \draw[very thick, red] (2,0) -- (2,-.6);
\end{tikzpicture} \hspace{15pt} = -\frac{1}{\lambda}\cdot \frac{1}{\lambda} \cdot 1
\end{align}
In this case there is no linking, because the blue chord can be slid to the left where it annihilates itself on the ``fold'' without crossing anything. 

But if we reverse the ordering of these two chords, then the red chord becomes an obstruction to sliding the blue chord to the left. This means that there is a nontrivial linking, and this contributes a factor of $q$ to the diagram:
\begin{align}
\begin{tikzpicture}[baseline={([yshift=-0.1cm]current bounding box.center)}]
    \draw[thick](0,0) -- (3,0) arc (90:-90:.15) -- (0,-.3) arc (90:270:.15) -- (3,-.6) arc (90:-90:.15) -- (0,-.9) arc (270:90:.45);
    \draw[very thick, blue] (2,-.3) -- (2,-.6);
    \draw[very thick, red] (1,0) -- (1,-.6);
\end{tikzpicture} \hspace{15pt} = -\frac{1}{\lambda}\cdot \frac{1}{\lambda} \cdot q
\end{align}

To study the more general problem, we can use an auxiliary vector space with vectors $|n\rangle$ that represent the case with $n$ obstructions to the left. Then acting with certain chords can increase the number of obstructions, by acting with an operator $\alpha^\dagger$ that raises the number by one $\alpha^\dagger |n\rangle = |n+1\rangle$. We have
\begin{align}
\begin{tikzpicture}[baseline={([yshift=-0.1cm]current bounding box.center)}]
    \draw[thick](0,0) -- (1,0);
    \draw[thick](0,-.3) -- (1,-.3);
    \draw[thick](0,-.6) -- (1,-.6);
    \draw[thick](0,-.9) -- (1,-.9);
    \draw[very thick] (.5,0) -- (.5,-.3);
\end{tikzpicture}
+
\begin{tikzpicture}[baseline={([yshift=-0.1cm]current bounding box.center)}]
    \draw[thick](0,0) -- (1,0);
    \draw[thick](0,-.3) -- (1,-.3);
    \draw[thick](0,-.6) -- (1,-.6);
    \draw[thick](0,-.9) -- (1,-.9);
    \draw[very thick] (.5,-.6) -- (.5,-.9);
\end{tikzpicture}
\hspace{10pt} &= \hspace{10pt} \frac{2}{\lambda}\alpha^\dagger \\[1em]
\begin{tikzpicture}[baseline={([yshift=-0.1cm]current bounding box.center)}]
    \draw[thick](0,0) -- (1,0);
    \draw[thick](0,-.3) -- (1,-.3);
    \draw[thick](0,-.6) -- (1,-.6);
    \draw[thick](0,-.9) -- (1,-.9);
    \draw[very thick] (.5,0) -- (.5,-.9);
\end{tikzpicture}
+
\begin{tikzpicture}[baseline={([yshift=-0.1cm]current bounding box.center)}]
    \draw[thick](0,0) -- (1,0);
    \draw[thick](0,-.3) -- (1,-.3);
    \draw[thick](0,-.6) -- (1,-.6);
    \draw[thick](0,-.9) -- (1,-.9);
    \draw[very thick] (.5,-.3) -- (.5,-.6);
\end{tikzpicture}
\hspace{10pt}&= \hspace{10pt} \frac{2}{\lambda} q^n \\[1em]
\begin{tikzpicture}[baseline={([yshift=-0.1cm]current bounding box.center)}]
    \draw[thick](0,0) -- (1,0);
    \draw[thick](0,-.3) -- (1,-.3);
    \draw[thick](0,-.6) -- (1,-.6);
    \draw[thick](0,-.9) -- (1,-.9);
    \draw[very thick] (.5,0) -- (.5,-.6);
\end{tikzpicture}
+
\begin{tikzpicture}[baseline={([yshift=-0.1cm]current bounding box.center)}]
    \draw[thick](0,0) -- (1,0);
    \draw[thick](0,-.3) -- (1,-.3);
    \draw[thick](0,-.6) -- (1,-.6);
    \draw[thick](0,-.9) -- (1,-.9);
    \draw[very thick] (.5,-.3) -- (.5,-.9);
\end{tikzpicture}
\hspace{10pt} &= \hspace{10pt} -\frac{2}{\lambda}\alpha^\dagger q^n \\[1em]
\begin{tikzpicture}[baseline={([yshift=-0.1cm]current bounding box.center)}]
    \draw[thick](0,0) -- (1,0);
    \draw[thick](0,-.3) -- (1,-.3);
    \draw[thick](0,-.6) -- (1,-.6);
    \draw[thick](0,-.9) -- (1,-.9);
    \draw[very thick] (.5,0) arc (180:360:.1);
\end{tikzpicture}
+
\begin{tikzpicture}[baseline={([yshift=-0.1cm]current bounding box.center)}]
    \draw[thick](0,0) -- (1,0);
    \draw[thick](0,-.3) -- (1,-.3);
    \draw[thick](0,-.6) -- (1,-.6);
    \draw[thick](0,-.9) -- (1,-.9);
    \draw[very thick] (.5,-.3) arc (180:360:.1);
\end{tikzpicture}
+
\begin{tikzpicture}[baseline={([yshift=-0.1cm]current bounding box.center)}]
    \draw[thick](0,0) -- (1,0);
    \draw[thick](0,-.3) -- (1,-.3);
    \draw[thick](0,-.6) -- (1,-.6);
    \draw[thick](0,-.9) -- (1,-.9);
    \draw[very thick] (.5,-.6) arc (180:360:.1);
\end{tikzpicture}
+
\begin{tikzpicture}[baseline={([yshift=-0.1cm]current bounding box.center)}]
    \draw[thick](0,0) -- (1,0);
    \draw[thick](0,-.3) -- (1,-.3);
    \draw[thick](0,-.6) -- (1,-.6);
    \draw[thick](0,-.9) -- (1,-.9);
    \draw[very thick] (.5,-.9) arc (180:360:.1);
\end{tikzpicture}
\hspace{10pt} &= \hspace{10pt}  -\frac{2}{\lambda}
\end{align}
Then we find
\begin{align}
\langle 1\rangle &=  \sum_{\text{chords}}\begin{tikzpicture}[baseline={([yshift=-0.1cm]current bounding box.center)}]
    \draw[thick](0,0) -- (3,0) arc (90:-90:.15) -- (0,-.3) arc (90:270:.15) -- (3,-.6) arc (90:-90:.15) -- (0,-.9) arc (270:90:.45);
\end{tikzpicture}\\
&=\sum_{n = 0}^\infty p(n).
\end{align}
where
\begin{align}
\sum_{n = 0}^\infty p(n)|n\rangle &=\exp\left\{\frac{2t}{\lambda}\left[ -1 - \alpha^\dagger q^n + \alpha^\dagger + q^n\right]\right\} |0\rangle \\
&= |0\rangle.
\end{align}
So we find that the sum over chords cancels out and $\langle 1\rangle = 1$ as it should.

Now we consider the case with matter operators inserted in the OTOC configuration considered in main text \ref{wvwvh}
\begin{align}
\langle e^{\Delta_1 g(0,0)}e^{\Delta_2 g(t,t)}\rangle_{\text{OTOC}} &= \sum_{\text{chords}}\begin{tikzpicture}[baseline={([yshift=-0.1cm]current bounding box.center)}]
    \draw[semithick, gray](0,0) -- (3,0) arc (90:-90:.15) -- (0,-.3) arc (90:270:.15) -- (3,-.6) arc (90:-90:.15) -- (0,-.9) arc (270:90:.45);
    \draw[very thick, black] (0,0) -- (3,0) arc (90:-90:.135) -- (0,-.275);
    \draw[very thick, black, dashed] (3,-.325) -- (0,-.325) arc (90:270:.135) -- (3,-.6);
    \filldraw[very thick, black, fill = white] (3,-.3) circle (.075);
    \filldraw[very thick, black, fill = white] (3,-.6) circle (.075);
    \filldraw[black] (0,0) circle (.075);
    \filldraw[black] (0,-.3) circle (.075);
\end{tikzpicture} \\
&= \sum_{m =0}^\infty p(m)
\end{align}
where
\begin{align}
\sum_{n = 0}^\infty p(n)|n\rangle =\exp\left[\frac{2t}{\lambda}(q^{\Delta_2}\alpha^\dagger -1)(1-q^{n+\Delta_1})\right] |0\rangle.\label{toDiag}
\end{align}

By using mathematica to do the exponentiation and series expansion to high enough order, we found
\begin{align}
\langle h(0,0)h(t,t)\rangle_{\text{OTOC}} = &-\lambda e^{2t} \\
&+ \frac{\lambda^2}{2}\left(e^{4t} - 2t e^{2t} -e^{2t}\right) \\
&-\frac{2\lambda^3}{3} \left(e^{6t}-3t e^{4t} - \frac{9}{8}e^{4t} + \frac{3}{4} t^2 e^{2t}+\frac{5}{4}t e^{2t} + \frac{1}{8} e^{2t}\right)\\
&+O(\lambda^4)
\end{align}
and
\begin{align}
\langle h(0,0)^2 h(t,t)^2\rangle &= 2\lambda^2 e^{4t}\\
&+ \lambda^3(-6 e^{6t} + 8 t e^{4t} + 7 e^{4t} - e^{2t})\\
&+\lambda^4(\frac{121}{6}e^{8t} -54 t e^{6t} - 30 e^{6t} + 16 t^2 e^{4t} + \frac{100}{3}t e^{4t} + \frac{31}{3}e^{4t} - t e^{2t} - \frac{1}{2}e^{2t})\\
&+O(\lambda^5).
\end{align}
We see consistency with the tree-level and one-loop computations in Section \ref{sec:two}, together with predictions for the two-loop corrections.

Now, let's try to compute (\ref{toDiag}) explicitly. We can start by trying to diagonalize the operator that appears there
\be
(q^{\Delta_2}\alpha^\dagger -1)(1-q^{n+\Delta_1})\sum_{n}\psi(n)|n\rangle = \lambda \sum_{n}\psi(n)|n\rangle
\ee
This gives
\be
\psi(n) = \frac{\lambda + 1-q^{1+n+\Delta_1}}{q^{\Delta_2}(1-q^{n+\Delta_1})}\psi(n+1).
\ee
The right eigenvectors $\psi_k(n)$ are labeled by some $k$ such that
\be
\psi_{k}(k-1) = 0, \hspace{20pt} \psi_k(k) = 1.
\ee
The corresponding eigenvalue is
\be
\lambda_k = -(1-q^{k+\Delta_1})
\ee
and the right eigenvector is
\be
\psi_k(k+m) = \begin{cases}0 & m < 0 \\
q^{m(\Delta_2-\Delta_1-k)}\frac{(q^{k+\Delta_1};q)_m}{(q;q)_m} & m \ge 0.
\end{cases}
\ee
The initial condition can be expressed in this eigenbasis as
\be
\delta_{n,1} = \sum_{k = 0}^\infty (-1)^kq^{k(\Delta_2-\Delta_1)-\frac{k(k-1)}{2}}\frac{(q^{\Delta_1};q)_k}{(q;q)_k}\psi_k(n).
\ee
Therefore the OTOC is 
\begin{align}
\langle e^{\Delta_1 g(0,0)} e^{\Delta_2 g(t,t)}\rangle_{\text{OTOC}} &=\sum_{n = 0}^\infty\sum_{k = 0}^n (-1)^kq^{k(\Delta_2-\Delta_1)-\frac{k(k-1)}{2}}\frac{(q^{\Delta_1};q)_k}{(q;q)_k}\times \psi_k(n)\times e^{-\frac{2t}{\lambda}(1-q^{k+\Delta_1})}\\
&= \sum_{n = 0}^\infty \sum_{k = 0}^n (-1)^k q^{n(\Delta_2-\Delta_1)-nk+k(k+1)/2}\frac{(q^{\Delta_1};q)_n}{(q;q)_k (q;q)_{n-k}} e^{-\frac{2t}{\lambda}(1-q^{k+\Delta_1})}\\
&=e^{-\frac{2t}{\lambda}}\sum_{n = 0}^\infty \frac{(q^{\Delta_1},q)_n}{(q,q)_n}q^{n(\Delta_2-\Delta_1)}\sum_{\ell = 0}^\infty \frac{\left(\frac{2t}{\lambda}q^{\Delta_1}\right)^\ell}{\ell!}(q^{1+\ell-n},q)_n.
\end{align}
In the last equation we used the $q$-Binomial theorem. Further simplifications may be possible.

\subsection{Propagator in a general configuration}\label{append:general}
We will now discuss the correlator $\lr h_{ij}(it_1,it_2)h_{i'j'}(it_3,it_4)\rr$. One can obtain it from
\begin{align}\label{twopointhh}
\langle h(it_1,it_2)h(it_3,it_4)\rangle= \partial_{\Delta}\partial_{\Delta'}\left(\langle e^{\Delta g(it_1,it_2)}e^{\Delta' g(it_3,it_4)}\rangle - \langle e^{\Delta g(it_1,it_2)}\rangle\langle e^{\Delta' g(it_3,it_4)}\rangle\right)\Big|_{\Delta = \Delta' = 0}.
\end{align}
(Here the contour indices has been suppressed.) In general the first expectation value has to be evaluated using a double Keldysh-Schwinger contour. But for certain operator configurations, it can be reduced to a single time fold, e.g.
\be
\begin{tikzpicture}[scale=0.7, rotate=0, baseline={([yshift=-0.15cm]current bounding box.center)}]
  \def\xc{5}
  \def\yy{.5}
  \def\yyy{.5}
  \draw[semithick, gray, rounded corners=.5mm] (0,0) -- (\xc,0) -- (\xc, -\yy) -- (0,-1*\yy) -- (0,-2.*\yy) -- (\xc, -2.*\yy) -- (\xc, -3.0*\yy) -- (0, -3.*\yy);
  \filldraw[black] (0,0) circle (0.1);
  \filldraw[black] (2*\xc/3,-3*\yy) circle (0.1);
  \filldraw[very thick, black, fill = white] (\xc/3,-1*\yy) circle (0.1);
  \filldraw[very thick, black, fill = white] (\xc,-3*\yy) circle (0.1);
\end{tikzpicture} \hspace{10pt}=\hspace{10pt} \begin{tikzpicture}[scale=0.75, rotate=0, baseline={([yshift=-0.15cm]current bounding box.center)}]
  \def\xc{5}
  \def\yy{.5}
  \draw[semithick, gray, rounded corners=.5mm] (0,0) -- (\xc,0) -- (\xc, -\yy) -- (0,-1*\yy);
%  \draw[very thick, black] (0,-0.04) -- (\xc-0.05,-0.04) -- (\xc-0.05, -\yy) -- (2*\xc/3,-1*\yy);
  %  \draw[very thick, black, dashed] (\xc/3,0.04) -- (\xc+0.05,0.04) -- (\xc+0.05, -\yy);
  \filldraw[black] (0,0) circle (0.1) node[anchor=south] {$t_1$};
  \filldraw[black] (2*\xc/3,-\yy) circle (0.1) node[anchor=north] {$t_2$};
  \filldraw[very thick, black, fill = white] (\xc/3,0) circle (0.1) node[anchor=south] {$t_3$};
  \filldraw[very thick, black, fill = white] (\xc,-\yy) circle (0.1) node[anchor=north] {$t_4$};
%  \node[text width = 60] at (\xc/2, -2*\yy) {\begin{center}$\lambda t$\end{center}};
 % \draw[|-|, very thick] (\xc/3,\yy) -- (2*\xc/3,\yy) node[midway, above] {$t$};
\end{tikzpicture}
\ee
On this single fold contour, both $\langle e^{\Delta g(it_1,it_2)}e^{\Delta' g(it_3,it_4)}\rangle$ and  $\langle e^{\Delta g(it_1,it_2)}\rangle\langle e^{\Delta' g(it_3,it_4)}\rangle$ can be computed using chords. For example, the computation of $\langle e^{\Delta g(it_1,it_2)}e^{\Delta' g(it_3,it_4)}\rangle$ involves the a matter chord extending from $t_1$ to $t_2$ and another matter chord extending from $t_3$ to $t_4$:
\be
\begin{tikzpicture}[scale=0.75, rotate=0, baseline={([yshift=-0.15cm]current bounding box.center)}]
  \def\xc{5}
  \def\yy{.5}
  \draw[semithick, gray, rounded corners=.5mm] (0,0) -- (\xc,0) -- (\xc, -\yy) -- (0,-1*\yy);
  \draw[very thick, black] (0,-0.04) -- (\xc-0.05,-0.04) -- (\xc-0.05, -\yy) -- (2*\xc/3,-1*\yy);
    \draw[very thick, black, dashed] (\xc/3,0.04) -- (\xc+0.05,0.04) -- (\xc+0.05, -\yy);
  \filldraw[black] (0,0) circle (0.1) node[anchor=south] {$t_1$};
  \filldraw[black] (2*\xc/3,-\yy) circle (0.1) node[anchor=north] {$t_2$};
  \filldraw[very thick, black, fill = white] (\xc/3,0) circle (0.1) node[anchor=south] {$t_3$};
  \filldraw[very thick, black, fill = white] (\xc,-\yy) circle (0.1) node[anchor=north] {$t_4$};
%  \node[text width = 60] at (\xc/2, -2*\yy) {\begin{center}$\lambda t$\end{center}};
 % \draw[|-|, very thick] (\xc/3,\yy) -- (2*\xc/3,\yy) node[midway, above] {$t$};
\end{tikzpicture}
\ee
We now consider the effect of Hamiltonian chords that connect the two sides of the fold. Between $t_1$ and $t_3$, such chords are only linked with black matter chord, giving a factor of $q^{\Delta}$ for each Hamiltonian chord. The sum over chords in this region gives
\be
 \sum_{m = 0}^\infty  \frac{(t_3-t_1)^m q^{\Delta m}}{m!}\frac{1}{\lambda^m} =e^{\frac{(t_3-t_1)q^{\Delta}}{\lambda}}
\ee
In the region between $t_3$ and $t_2$, we a factor of $q^{\Delta+\Delta'}$ for each Hamiltonian chord, and between $t_2$ and $t_4$ we get a factor of $q^{\Delta'}$. Including also the chords linking the same side, we get the exact final answer
\be
\begin{aligned}
&\langle e^{\Delta g(it_1,it_2)}e^{\Delta' g(it_3,it_4)}\rangle=e^{\frac{(t_3-t_1)q^{\Delta}}{\lambda}} e^{\frac{(t_2-t_3)q^{\Delta+\Delta'}}{\lambda}} e^{\frac{(t_4-t_2)q^{\Delta'}}{\lambda}} e^{-\frac{1}{\lambda}(t_4-t_1)}.
\end{aligned}
\ee
For $\langle e^{\Delta g(it_1,it_2)}\rangle\langle e^{\Delta' g(it_3,it_4)}\rangle$, this is a product over two 2-point functions computed in (\ref{2ptchord})
\be
\begin{aligned}
\langle e^{\Delta g(it_1,it_2)}\rangle\langle e^{\Delta' g(it_3,it_4)}\rangle=e^{\frac{(t_2-t_1)q^{\Delta}}{\lambda}} e^{\frac{(t_4-t_3)q^{\Delta'}}{\lambda}} e^{-\frac{1}{\lambda}(t_2-t_1)}e^{-\frac{1}{\lambda}(t_4-t_3)}.
\end{aligned}
\ee
Now plugging into (\ref{twopointhh}), we find the exact answer
\be
\langle h(it_1,it_2)h(it_3,it_4)\rangle=\lambda(t_2-t_3).
\ee

Having studied this case, let's now classify all possible configurations of $\lr h_{ij}(it_1,it_2)h_{i'j'}(it_3,it_4)\rr$. Without loss of generality, we take $t_1<t_2$ and $t_3<t_4$.  We first consider cases where $t_1,t_2$ are adjacent on the contour. This will automatically mean $t_3,t_4$ are adjacent. In this case, all configurations on a doubled Schwinger-Keldysh contour can be represented using a single time-fold:
\be
\begin{tikzpicture}[scale=0.75, rotate=0, baseline={([yshift=-0.15cm]current bounding box.center)}]
  \def\xc{5}
  \def\yy{.5}
  \draw[white] (0,2.375*\yy) -- (\xc,2.375*\yy);
  \draw[very thick, gray, rounded corners=.5mm] (0,0) -- (\xc,0) -- (\xc, -\yy) -- (0,-1*\yy);
  \filldraw[black] (0,0) circle (0.1);
  \filldraw[black] (\xc/3,0) circle (0.1);
  \filldraw[very thick, black, fill = white] (2*\xc/3,0) circle (0.1);
  \filldraw[very thick, black, fill = white] (\xc,0) circle (0.1);
  \node[text width = 60] at (\xc/2, -2*\yy) {\begin{center}$0$\end{center}};
\end{tikzpicture} \hspace{30pt}
\begin{tikzpicture}[scale=0.75, rotate=0, baseline={([yshift=-0.15cm]current bounding box.center)}]
  \def\xc{5}
  \def\yy{.5}
  \draw[very thick, gray, rounded corners=.5mm] (0,0) -- (\xc,0) -- (\xc, -\yy) -- (0,-1*\yy);
  \filldraw[black] (0,0) circle (0.1);
  \filldraw[black] (2*\xc/3,-\yy) circle (0.1);
  \filldraw[very thick, black, fill = white] (\xc/3,0) circle (0.1);
  \filldraw[very thick, black, fill = white] (\xc,-\yy) circle (0.1);
  \node[text width = 60] at (\xc/2, -2*\yy) {\begin{center}$\lambda t$\end{center}};
  \draw[|-|, very thick] (\xc/3,\yy) -- (2*\xc/3,\yy) node[midway, above] {$t$};
\end{tikzpicture} \hspace{30pt}
\begin{tikzpicture}[scale=0.75, rotate=0, baseline={([yshift=-0.15cm]current bounding box.center)}]
  \def\xc{5}
  \def\yy{.5}
  \draw[very thick, gray, rounded corners=.5mm] (0,0) -- (\xc,0) -- (\xc, -\yy) -- (0,-1*\yy);
  \filldraw[black] (0,0) circle (0.1);
  \filldraw[black] (3*\xc/3,0) circle (0.1);
  \filldraw[very thick, black, fill = white] (\xc/3,0) circle (0.1);
  \filldraw[very thick, black, fill = white] (2*\xc/3,0) circle (0.1);
  \node[text width = 60] at (\xc/2, -2*\yy) {\begin{center}$\lambda t$\end{center}};
  \draw[|-|, very thick] (\xc/3,\yy) -- (2*\xc/3,\yy) node[midway, above] {$t$};
\end{tikzpicture} 
\ee
We wrote the answer for $\langle h(it_1,it_2)h(it_3,it_4)\rangle$ below the corresponding diagram. The definition of $t$ is also labelled in the graph. The middle configuration is the example we calculated in detail above and the other cases are evaluated similarly.

Now let's consider diagrams where $t_1,t_2$ are not adjacent on the Schwinger-Keldysh contour:
\be
\begin{tikzpicture}[scale=0.75, rotate=0, baseline={([yshift=-0.15cm]current bounding box.center)}]
  \def\xc{5}
  \def\yy{.5}
  \draw[very thick, gray, rounded corners=.5mm] (0,0) -- (2*\xc/3,0) -- (2*\xc/3, -\yy) -- (\xc/3,-1*\yy) -- (\xc/3,-2.*\yy) -- (\xc, -2.*\yy) -- (\xc, -3.0*\yy) -- (0, -3.*\yy);
  \filldraw[black] (0,0) circle (0.1);
  \filldraw[black] (\xc/3,-2*\yy) circle (0.1);
  \filldraw[very thick, black, fill = white] (2*\xc/3,-1*\yy) circle (0.1);
  \filldraw[very thick, black, fill = white] (\xc,-3*\yy) circle (0.1);
  \node[text width = 100] at (\xc/2, -4*\yy) {\begin{center}$-\lambda e^{2t} + \dots$\end{center}};
  \draw[|-|, very thick] (\xc/3,\yy) -- (2*\xc/3,\yy) node[midway, above] {$t$};
\end{tikzpicture} \hspace{30pt}
\begin{tikzpicture}[scale=0.75, rotate=0, baseline={([yshift=-0.15cm]current bounding box.center)}]
  \def\xc{5}
  \def\yy{.5}
  \draw[very thick, gray, rounded corners=.5mm] (0,0) -- (\xc,0) -- (\xc, -\yy) -- (0,-1*\yy);
  \filldraw[black] (0,0) circle (0.1);
  \filldraw[black] (2*\xc/3,0) circle (0.1);
  \filldraw[very thick, black, fill = white] (\xc/3,0) circle (0.1);
  \filldraw[very thick, black, fill = white] (\xc,0) circle (0.1);
  \node[text width = 60] at (\xc/2, -4*\yy) {\begin{center}$\lambda (t-1)$\end{center}};
  \draw[|-|, very thick] (\xc/3,\yy) -- (2*\xc/3,\yy) node[midway, above] {$t$};
\end{tikzpicture} \hspace{30pt}
\begin{tikzpicture}[scale=0.75, rotate=0, baseline={([yshift=-0.15cm]current bounding box.center)}]
  \def\xc{5}
  \def\yy{.5}
  \draw[very thick, gray, rounded corners=.5mm] (0,0) -- (\xc,0) -- (\xc, -\yy) -- (0,-1*\yy);
  \filldraw[black] (0,0) circle (0.1);
  \filldraw[black] (\xc,0) circle (0.1);
  \filldraw[very thick, black, fill = white] (\xc/3,0) circle (0.1);
  \filldraw[very thick, black, fill = white] (2*\xc/3,-1*\yy) circle (0.1);
  \node[text width = 60] at (\xc/2, -4*\yy) {\begin{center}$\lambda (t-1)$\end{center}};
  \draw[|-|, very thick] (\xc/3,\yy) -- (2*\xc/3,\yy) node[midway, above] {$t$};
\end{tikzpicture} 
\ee
The first case was analyzed in the main text; to compute it using chords one can use the method in section \ref{usemethod}. Note that there are a series of $\lambda^k$ corrections in this case. The other two cases have simple exact formulas, and the computation differs from the previous cases only due to a factor of $q^{\Delta\Delta'}$ for the crossing of the matter chords themselves.

\section{Two-site large \texorpdfstring{$p$}{p} model}

In this appendix, we consider the special case of the large $p$ Brownian SYK chain \eqref{1daction} with two sites. Let us label the two sites $A$ and $B$. The Liouville action for this case is 
\begin{align} 
&I = I_0 + I_{\rm int} \, ,  \\
&I_0  = \sum_{i,j} \sigma(i, j)   \sum_{x=A,B}\frac{1}{2\lambda} \int_0^t \d t_1 \int_0^t \d t_2 \bigg[ - \frac{1}{4} g_{x, ij}  \partial_1 \partial_2 g_{x, ij} + e^{g_{x, ij}(\i t_1, \i t_2)} \delta(t_1-t_2)\bigg] \\
& I_{\rm int}  =  \sum_{i, j} \sigma(i,j) \left[ - \sum_{x=A,B} \frac{a}{2\lambda}  \int_0^t \d t_1 e^{g_{x, ij}(\i t_1, \i t_1)} + \frac{a}{\lambda} \int_0^t \d t_1  e^{\frac{g_{A, ij}(\i t_1,\i t_1) + g_{B, ij}(\i t_1, \i t_1)}{2}} \ri]\, . \label{iint}
\end{align} 
Below $\braket{...}$ will refer to expectation values in the full action $I$, and $\braket{...}_0$ will refer to expectation values in $I_0$.
 The saddle-point equations and saddle-point value of the action are both the same for $I_0$ and $I$, so in particular the saddle-point value of $g_{x, ij}(\i t, \i t)$ is zero, and  
\be 
\braket{e^{-I_{\rm int}}}_{0} = 1\, . 
\ee
Let us again introduce the notation $\h_{ij, x}(t)$ for the fluctuations of $g_{x, ij}(\i t, \i t)$ around the saddle-point value as in the main text. 
 We will evaluate the OTOC 
\be 
\braket{W_A(0) V_B(t) W_A(0) V_B(t)}= \braket{e^{\Delta_A  \, \h_{A, 12}(0)} e^{\Delta_B \,  \h_{B, 23}(t)} }  \, , 
\ee
where $W_A$ and $V_B$ are operators of the kind defined in \eqref{wdef} at sites $A$ and $B$, and are represented by solid or empty circles below:
\begin{figure}[!h]
\centering
\begin{tikzpicture}[baseline={([yshift=-0.1cm]current bounding box.center)}]
    \draw[thick](0,0) -- (3,0) arc (90:-90:.15) -- (0,-.3) arc (90:270:.15) -- (3,-.6) arc (90:-90:.15) -- (0,-.9) arc (270:90:.45);
   % \draw[very thick, blue] (0,0) -- (3,0) arc (90:-90:.135) -- (0,-.275);
   % \draw[very thick, red] (3,-.325) -- (0,-.325) arc (90:270:.135) -- (3,-.6);
    \filldraw[very thick, black, fill = white] (3,-.3) circle (.075);
    \filldraw[very thick, black, fill = white] (3,-.6) circle (.075);
    \filldraw[black] (0,0) circle (.075);
    \filldraw[black] (0,-.3) circle (.075);
\end{tikzpicture}
%\caption{}
%\label{fig:ABSK} 
\end{figure} 

We will be interested in times late enough such that each of the single sites is fully scrambled. Let us first consider a regime where $a \ll 1$ and treat $I_{\rm int}$ as a perturbation. 
We then have 
\begin{align}
&\braket{e^{\Delta_A \, \h_{A, 12}(0)} e^{\Delta_B \, \h_{B, 23}(t)} }  = \frac{1}{\braket{e^{-I_{\rm int}}}}_0 \braket{e^{- I_{\rm int}}  e^{\Delta_A \h_{A, 12}(0)} e^{\Delta_B \h_{B, 23}(t)} }_0 \label{16} \\ 
%& =  \braket{e^{\Delta_A \, g_{A, 12}(t,t)}}_0\braket{e^{\Delta_B \,  g_{B, 23}(0,0)}}_0  + ... \\
 &=  \braket{e^{\Delta_A \, \h_{12}(0)}}_0\braket{e^{\Delta_B \,  \h_{23}(t)}}_0  \\
& + \frac{a}{2\lambda} \sum_{i,j} \sigma(i,j) \int_0^t dt_1 \braket{e^{\Delta_A \,  \h_{12}(0)} e^{\h_{ij}(t_1)}}_0 \braket{e^{\Delta_B \, \h_{23}(t)}}_0 \label{17}\\
& +\frac{a}{2\lambda} \sum_{i,j} \sigma(i,j) \int_0^t dt_1 \braket{ e^{\h_{ij}(t_1)}e^{\Delta_B \, \h_{23}(t)}}_0 \braket{e^{\Delta_A\,  \h_{23}(0)}}_0  \label{18} \\
&- \frac{a}{\lambda} \sum_{i,j} \sigma(i,j) \int_0^t dt_1 \braket{e^{\Delta_A \, \h_{12}(0)} e^{\ha \h_{ij}(t_1)}}_0 \braket{e^{\ha \h_{ij}(t_1)}e^{\Delta_B \, \h_{23}(t)} }_0  \label{19}  \\&+  O(a^2)
\end{align} 
In the final expression, since all expectation values are in $I_0$, we have removed the $A$ and $B$ labels from the $g_{ij}$ fields, and the 0 subscript denotes expectation values in a single-site SYK model. Recall that 
\begin{align} 
&\braket{e^{\Delta \h_{ij}(t)}}_0 = 1,  \label{g1}\\
&\braket{e^{\Delta_1 \h_{i_1j_1}(t_1)} e^{\Delta_2 \h_{i_2j_2}(t_2)} }_0= 1 \quad \text{ for time-ordered configurations } i_1, j_1, i_2, j_2 \, . \label{g2}
\end{align} 
%\SV{[\eqref{g2} is based on the fact that we see this from the early time Brownian calculation to all orders in $1/p$?]}
Using \eqref{g1} and \eqref{g2}, we can check that the terms \eqref{17} and \eqref{18} are both zero due to the sum over contours. This makes sense, as these contributions come from the terms in $I_{\rm int}$ which do not involve interactions between $A$ and $B$. Let us consider the various contour choices $\{i, j\}$ in the final term. Let us denote the OTOC for the single-site model as follows: 
\be
F_{\Delta,\Delta'}(t) = \langle e^{\Delta \h(0)}e^{\Delta' \h(t)}\rangle_{\rm OTOC, \, 0}\, . 
\ee
Then we get the following contributions to \eqref{19}: 
 \begin{align}
&\{1,1\} \hspace{20pt}\implies \hspace{20pt} \begin{tikzpicture}[baseline={([yshift=-0.1cm]current bounding box.center)}]
    \draw[thick](0,0) -- (3,0) arc (90:-90:.15) -- (0,-.3) arc (90:270:.15) -- (3,-.6) arc (90:-90:.15) -- (0,-.9) arc (270:90:.45);
   % \draw[very thick, blue] (0,0) -- (3,0) arc (90:-90:.135) -- (0,-.275);
   % \draw[very thick, red] (3,-.325) -- (0,-.325) arc (90:270:.135) -- (3,-.6);
    \filldraw[very thick, black, fill = white] (3,-.3) circle (.075);
    \filldraw[very thick, black, fill = white] (3,-.6) circle (.075);
    \filldraw[black] (0,0) circle (.075);
    \filldraw[black] (0,-.3) circle (.075);
    \filldraw[red] (1,0) circle (.075);
    \filldraw[red] (1.15,0) circle (.075);
\end{tikzpicture} \hspace{20pt} \implies \hspace{20pt} -\frac{a}{\lambda}\int_0^t d t_1\, {1}\cdot { 1}\\
&\{1,2\} \hspace{20pt}\implies \hspace{20pt} \begin{tikzpicture}[baseline={([yshift=-0.1cm]current bounding box.center)}]
    \draw[thick](0,0) -- (3,0) arc (90:-90:.15) -- (0,-.3) arc (90:270:.15) -- (3,-.6) arc (90:-90:.15) -- (0,-.9) arc (270:90:.45);
   % \draw[very thick, blue] (0,0) -- (3,0) arc (90:-90:.135) -- (0,-.275);
   % \draw[very thick, red] (3,-.325) -- (0,-.325) arc (90:270:.135) -- (3,-.6);
    \filldraw[very thick, black, fill = white] (3,-.3) circle (.075);
    \filldraw[very thick, black, fill = white] (3,-.6) circle (.075);
    \filldraw[black] (0,0) circle (.075);
    \filldraw[black] (0,-.3) circle (.075);
    \filldraw[red] (1,0) circle (.075);
    \filldraw[red] (1,-.3) circle (.075);
\end{tikzpicture} \hspace{20pt} \implies \hspace{20pt} +\frac{a}{\lambda}\int_0^t d t_1 \,{ 1}\cdot {F_{\Delta_B,\frac{1}{2}}(t-t_1)}\\
&\{1,3\} \hspace{20pt}\implies \hspace{20pt} \begin{tikzpicture}[baseline={([yshift=-0.1cm]current bounding box.center)}]
    \draw[thick](0,0) -- (3,0) arc (90:-90:.15) -- (0,-.3) arc (90:270:.15) -- (3,-.6) arc (90:-90:.15) -- (0,-.9) arc (270:90:.45);
   % \draw[very thick, blue] (0,0) -- (3,0) arc (90:-90:.135) -- (0,-.275);
   % \draw[very thick, red] (3,-.325) -- (0,-.325) arc (90:270:.135) -- (3,-.6);
    \filldraw[very thick, black, fill = white] (3,-.3) circle (.075);
    \filldraw[very thick, black, fill = white] (3,-.6) circle (.075);
    \filldraw[black] (0,0) circle (.075);
    \filldraw[black] (0,-.3) circle (.075);
    \filldraw[red] (1,0) circle (.075);
    \filldraw[red] (1,-.6) circle (.075);
\end{tikzpicture} \hspace{20pt} \implies \hspace{20pt} -\frac{a}{\lambda}\int_0^t d t_1 {F_{\Delta_A,\frac{1}{2}}(t_1)}\cdot{F_{\Delta_B,\frac{1}{2}}(t-t_1)}\notag \\
&\{1,4\} \hspace{20pt}\implies \hspace{20pt} \begin{tikzpicture}[baseline={([yshift=-0.1cm]current bounding box.center)}]
    \draw[thick](0,0) -- (3,0) arc (90:-90:.15) -- (0,-.3) arc (90:270:.15) -- (3,-.6) arc (90:-90:.15) -- (0,-.9) arc (270:90:.45);
   % \draw[very thick, blue] (0,0) -- (3,0) arc (90:-90:.135) -- (0,-.275);
   % \draw[very thick, red] (3,-.325) -- (0,-.325) arc (90:270:.135) -- (3,-.6);
    \filldraw[very thick, black, fill = white] (3,-.3) circle (.075);
    \filldraw[very thick, black, fill = white] (3,-.6) circle (.075);
    \filldraw[black] (0,0) circle (.075);
    \filldraw[black] (0,-.3) circle (.075);
    \filldraw[red] (1,0) circle (.075);
    \filldraw[red] (1,-.9) circle (.075);
\end{tikzpicture} \hspace{20pt} \implies \hspace{20pt} +\frac{a}{\lambda}\int_0^t dt_1 {F_{\Delta_A,\frac{1}{2}}(t_1)} \cdot 1
\end{align}
The remaining choices of $\{i,j\}$ contribute a factor of four, so we find
\be
\braket{e^{\Delta_A \, \h_{A}(0)} e^{\Delta_B \, \h_{B}(t)} }_{\rm OTOC} = 1 - \frac{4a}{\lambda} \int_0^t d t_1 \Big[1-F_{\Delta_A,\frac{1}{2}}(t_1)\Big]\Big[1-F_{\Delta_B,\frac{1}{2}}(t-t_1)\Big] + O(a^2). \label{119}
\ee
Let us evaluate this in the case where $t$ is very large, $t \gg \log \frac{1}{\lambda}$. Then for most of the integration range between zero and $t$, the OTOC functions will have decayed, and we can approximate the integrand as one.
%~\footnote{We get the condition $t \gg \log \frac{1}{\lambda}$ from requiring the contribution to the integral $\int_0^t d t_1F_{\Delta_A,\frac{1}{2}}(t_1)F_{\Delta_B,\frac{1}{2}}(t-t_1)$ from $t_1 \sim t/2$ to be small.} 
So in this limit, we get
\be
\langle e^{\Delta_A \h(0)}e^{\Delta_B \h(t)}\rangle_{\rm OTOC} \approx 1 - \frac{4at}{\lambda}+ O(a^2). \label{120}
\ee

Let us now consider higher orders in $a$. Again we expect all contributions from the first term of \eqref{iint} to cancel and give no contribution to this OTOC, so let us simply remove this term from the action.  Then from \eqref{16}, we have
\begin{align} 
&\braket{e^{\Delta_A \, \h_{A, 12}(0)} e^{\Delta_B \, \h_{B, 23}(t)} } \nn
&= \sum_{n=0}^{\infty} \frac{1}{n!} \bigg \langle e^{\Delta_A \, \h_{A, 12}(0)} e^{\Delta_B \, \h_{B, 23}(t)}\left[-\sum_{i,j}\sigma(i,j) \frac{a}{\lambda} \int_0^t dt_1 e^{(\h_{A, ij}(t_1) +\h_{B, ij}(t_1))/2}\ri]^n\bigg \rangle_0 
\end{align} 
By summing the terms for any $n$ where all factors coming from the square bracket have $i=j$, we get a contribution $e^{-\frac{4at}{\lambda}}$. 
All other terms involve higher-point out-of-time-ordered correlators, and decay for sufficiently late times. 
We therefore have 
\be
\langle e^{\Delta_A \h(0)}e^{\Delta_B \h(t)}\rangle_{\rm OTOC} \approx  e^{-\frac{4at}{\lambda}} \, . \label{exp_a}
\ee

To better understand these results, let us consider the expansion of \eqref{119} in powers of $\lambda$. 
Using the multi-scramblon resummation result for $F_{\Delta,\Delta'}(t)$ from  \eqref{eikonalresum} in \eqref{119}, we get  
\be
\langle e^{\Delta_A \h_A(0)}e^{\Delta_B \h_B(t)}\rangle_{\rm OTOC}= 1 -\frac{4a}{\lambda}\sum_{m,n=1}^\infty \frac{(\Delta_A)_n (\frac{1}{2})_n}{n!}\frac{(\Delta_B)_m (\frac{1}{2})_m}{m!}(-\lambda)^{n+m}\int_0^t d t' e^{2nt'}e^{2m(t-t')} \label{126}
\ee
 The $m, n$ term in the above sum gives a contribution proportional to 
\be
\lambda^{n+m-1} e^{2\text{max}(n,m)t} \, .  \label{mn}
\ee
Recall that we are in the regime where $1 \ll \lambda e^{t} \ll \lambda e^{2t}$, $\lambda \ll 1$. To a first approximation, we might consider keeping only terms of the form $(\lambda e^{2t})^p$ for some integer $p$ in \eqref{126}, and ignoring terms of the form $\lambda^p e^{2tq}$ for $p>q$. From \eqref{mn}, all terms of the form $(\lambda e^{2t})^p$  come from the case where either $m=1$ or $n=1$ or both. However, on keeping only such terms, we get the following expression for the OTOC: 
\be
\langle e^{\Delta_A \h_A(0)}e^{\Delta_B \h_B(t)}\rangle_{\rm OTOC} \approx  1 - a(\Delta_A+\Delta_B) e^{2t}
\ee
which is of a completely different qualitative form from the result \eqref{120}. We therefore need to keep all terms in the sum over $m, n$ in \eqref{126}.

\bibliography{references}

\bibliographystyle{utphys}

\end{document}